\documentclass[reprint,twocolumn, pra, showpacs, superscriptaddress, floatfix, longbibliography]{revtex4-2}
\usepackage{bm}
\usepackage{xcolor}
\usepackage{hyperref}
\usepackage{graphicx}
\usepackage{float}
\usepackage[normalem]{ulem}
\usepackage[english]{babel}
\usepackage{algpseudocode}
\usepackage{amsmath}
\usepackage{amssymb}
\usepackage{bbold}
\usepackage{comment}
\usepackage{subfiles}
\usepackage[ruled,vlined]{algorithm2e}
\usepackage{ragged2e}

\makeatletter
\newcommand{\graphicalabstract}{%

\onecolumngrid

\begin{center}

\begin{minipage}{0.85\textwidth}
\centering

\includegraphics[width=\textwidth]{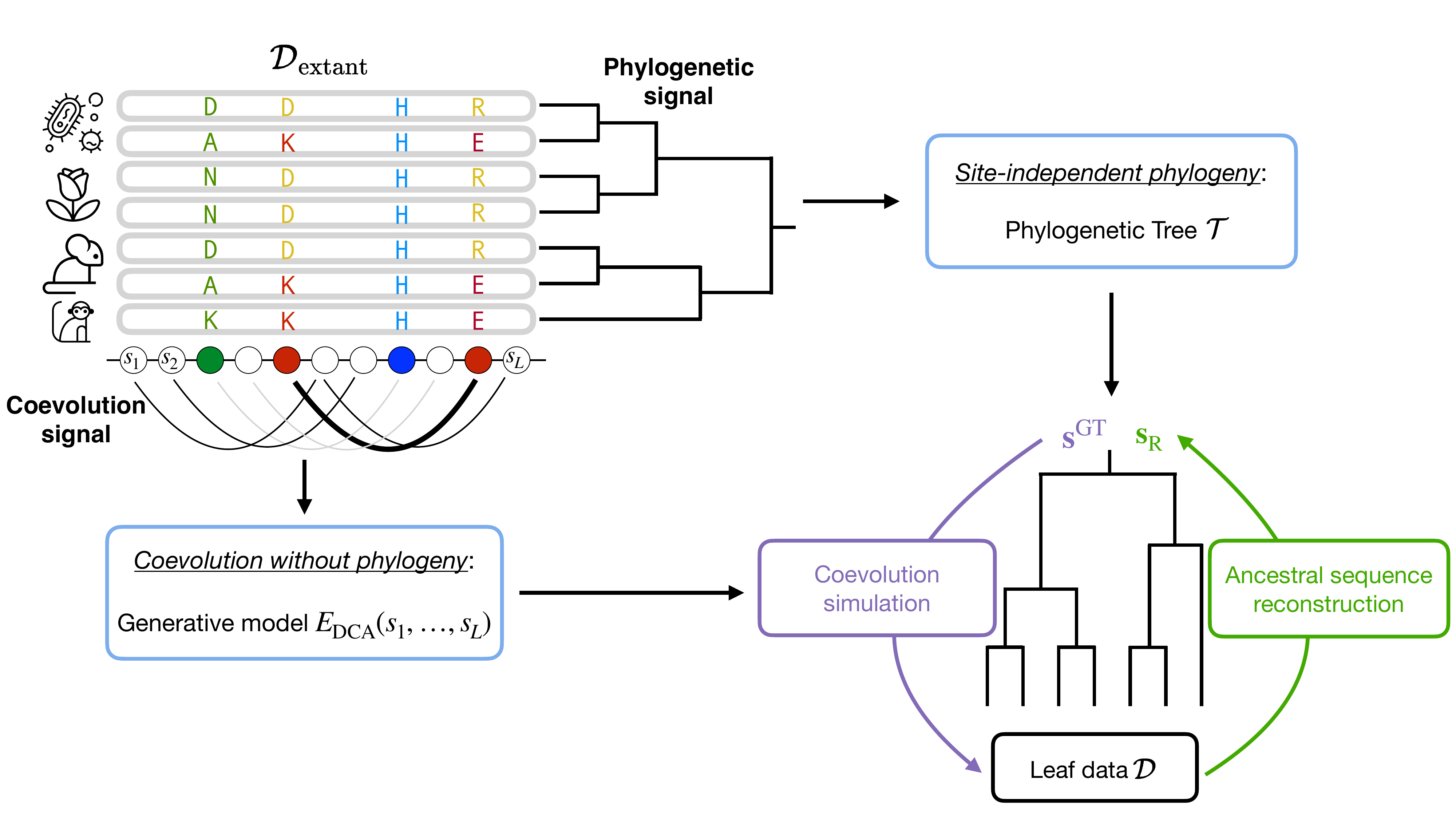}

\vspace{0.3em}

\small
\justifying
\textbf{Graphical abstract.} Our procedure works as follows: we take as input a Multiple Sequence Alignment of extant sequences $\Data_\text{extant}$, and infer in parallel both a phylogenetic tree $\Tree$ (phylogenetic signal) and a Direct Coupling Analysis model of coevolution (generative model with energy $E_\text{DCA}$). The two models are then combined to form a general, coevolution-aware framework for Ancestral Sequence Reconstruction, which can be benchmarked against \textit{in silico} data generated by the DCA forward evolver.

\end{minipage}

\end{center}

\clearpage
\twocolumngrid

}
\makeatother

\newcommand{\vect}[1]{\bm{#1}}
\newcommand{\postpersite}{\mathbb{P}^{(i)}_\text{R}}

\newcommand{\Data}{\vect{\mathcal{D}}}
\newcommand{\DataVec}[1]{\vect{\mathcal{D}}_{#1}}
\newcommand{\Tree}{\mathcal{T}}
\newcommand{\Alphabet}{\mathcal{A}}
\newcommand{\ancestralset}{\vect{\mathcal{S}}_\text{anc}}
\newcommand{\DCAset}{\vect{\mathcal{S}}_\text{anc}^\text{DCA}}

\newcommand{\LikRoot}{\mathcal{L}_\text{R}}

\newcommand{\DCAprob}{\mathcal{P}_\text{DCA}}

\newcommand{\Post}{\mathbb{P}_\text{R}}

\newcommand{\muDCA}{\mu_\text{gen}}

\newcommand{\Seq}{\vect{s}}

\newcommand{\RootSeq}{\vect{s}_\text{R}}

\newcommand{\GT}{\vect{s}^\text{GT}}
\newcommand{\GTindexed}[1]{\vect{s}^\text{GT\,#1}}

\newcommand{\Cons}{\vect{s}^{\text{cons}}}

\newcommand{\MAP}{\vect{s}^{\text{MAP}}}
\newcommand{\MAPsite}[1]{s^{\text{MAP}}_{#1}}

\newcommand{\given}{\,\mid\,}

\hypersetup{
    colorlinks=true,
    linkcolor=blue,
    citecolor=blue,
    urlcolor=blue,
}

\usepackage{etoolbox}
\robustify{\vect}
\robustify{\GT}
\robustify{\GTindexed}

\def\to{\rightarrow}  

\newcommand{\beq}{\begin{equation}} \newcommand{\eeq}{\end{equation}}

\begin{document}



\title{Towards coevolution-aware ancestral sequence reconstruction
}

\author{Alya Zeinaty}%
\affiliation{Sorbonne Universit\'e, CNRS, Computational, Quantitative and Synthetic Biology, 75005 Paris, France}
\affiliation{Dipartimento di Fisica, Sapienza Universit\`a di Roma, Piazzale Aldo Moro 5, 00185 Rome, Italy}
\author{Leonardo di Bari}%
\affiliation{Sorbonne Universit\'e, CNRS, Computational, Quantitative and Synthetic Biology, 75005 Paris, France}
\affiliation{DISAT, Politecnico di Torino, Corso Duca degli Abruzzi, 24, I-10129, Torino, Italy}
\author{Saverio Rossi}%
\affiliation{Dipartimento di Fisica, Sapienza Universit\`a di Roma, Piazzale Aldo Moro 5, 00185 Rome, Italy}
\author{Pierre~Barrat-Charlaix}%
\affiliation{Sorbonne Universit\'e, CNRS, Computational, Quantitative and Synthetic Biology, 75005 Paris, France}
\author{Francesco Zamponi}\thanks{Corresponding authors. Email: francesco.zamponi@uniroma1.it, martin.weigt@upmc.fr}%
\affiliation{Dipartimento di Fisica, Sapienza Universit\`a di Roma, Piazzale Aldo Moro 5, 00185 Rome, Italy}
\author{Martin Weigt}\thanks{Corresponding authors. Email: francesco.zamponi@uniroma1.it, martin.weigt@upmc.fr}%
\affiliation{Sorbonne Universit\'e, CNRS, Computational, Quantitative and Synthetic Biology, 75005 Paris, France}
\affiliation{Institut Universitaire de France (IUF)}


\begin{abstract}
Ancestral sequence reconstruction (ASR) is a powerful approach for studying molecular evolution and the emergence of protein function. Yet most ASR methods assume that sites evolve independently, neglecting the epistatic constraints that shape protein structure, stability, and function. This simplification affects both ancestral inference and its evaluation: maximum-a-posteriori reconstructions may over-concentrate probability into a single over-idealized sequence, whereas independent posterior sampling can generate implausible or poorly functional ancestors. Here, we introduce a coevolution-aware ASR framework that combines standard phylogenetic inference with Direct Coupling Analysis (DCA), thereby preserving site-wise ancestral uncertainty while enforcing residue–residue constraints learned from extant protein families. To benchmark the method, we develop a controlled forward-evolution framework based on a DCA evolutionary sampler, allowing reconstructed ancestors to be compared with known ground-truth sequences generated under realistic epistatic constraints. Applied to $\beta$-lactamases and DNA-binding domains, the approach improves reconstruction when ancestral states are epistatically constrained, and yields ensembles of candidate ancestors that are both phylogenetically consistent and statistically compatible with natural protein families. This framework bridges the gap between single-sequence MAP reconstruction and unconstrained posterior sampling, providing a practical route toward ancestral reconstructions that better reflect the coupled nature of protein evolution.
\end{abstract}
\setlength{\abovecaptionskip}{4pt}
\setlength{\belowcaptionskip}{4pt}

\maketitle
\graphicalabstract

\section{Introduction}

Ancestral sequence reconstruction (ASR) was first proposed as a means to infer ancestral phenotypes, and thus gain insight into the mechanisms of molecular evolution~\cite{pauling_1963}. 
Since then, ASR has become a central tool in evolutionary biology~\cite{liberles2007ancestral}, enabling the study of sequence–function relationships across deep evolutionary timescales~\cite{harms2010analyzing}. 
ASR also leads to experimental resurrection of functional ancestral proteins with novel or enhanced properties~\cite{thornton2004resurrecting,prakineeAncestralSequenceReconstruction2024}, as well as to the optimization of proteins' pharmaceutical properties~\cite{zakas2017enhancing}. 
As a result, ASR has been and remains a critical method to uncover novel insights in evolutionary processes and function emergence. 

Current ASR methods are based on models in which each site (nucleotide or amino acid) evolves independently of the others, and mainly use the Yang algorithm~\cite{yangNewMethodInference1995} which extends the Felsenstein tree-pruning algorithm for reconstruction of internal node states~\cite{felsensteinEvolutionaryTreesDNA1981}. 
These models generally maintain homogeneity of evolutionary parameters across the phylogeny, though they may incorporate site-specific variation in substitution rate and equilibrium frequencies. As a result, such site-independent models cannot accurately represent the important structural and functional constraints that influence sequence evolution~\cite{cocco2018inverse}.

Within this framework, ancestral sequences are typically inferred through Bayesian inference, which provides a posterior distribution $\postpersite(\cdot \given \DataVec{i}, \Tree, \vect{\theta}_i)$ of possible amino acids for each residue $i \in \{1,\cdots, L\}$, with $L$ the length of the sequence.
at the ancestral node $\text{R}$, given data at the leaves $\Data$, phylogenetic tree $\Tree$, and parameters $\vect{\theta}$ of the time-reversible, site-independent evolutionary model.
Most studies~\cite{zhaoAncestralSequenceReconstruction2026,chernyavskayaAncestralIntronicSplicing2026, prakineeAncestralSequenceReconstruction2024, supekarAncestralSequenceReconstruction2026, spenceAncestralSequenceReconstruction2021} focus on the maximum-a-posteriori (MAP) estimate $\MAP$, which assigns to each site the state with the largest posterior probability as 

\begin{equation}\label{eq:MAP}
    \MAPsite{i} = \arg\max_{a \in \Alphabet}
\postpersite(a\given \DataVec{i}, \Tree, \vect{\theta}_i),
\end{equation}
with $\Alphabet$ the alphabet of possible characters (amino acids or nucleotides), and index $i$ indicating data and parameters reduced to position $i$.  
MAP estimates in ASR have been found to exhibit exceptional properties, such as unusually high thermo-stability and catalytic promiscuity compared to extant proteins~\cite{spenceAncestralSequenceReconstruction2021}, making them ideal reconstructed sequences.
On the other hand, few works used the Bayesian approach of sampling ancestors directly from the site-independent posterior $\Post$, partly due to the experimental cost of testing multiple sequences, and partly because Bayesian sampling sometimes results in non-functional sequences, especially when the reconstruction was uncertain to begin with~\cite{eickRobustnessReconstructedAncestral2017}. 

Most importantly, these site-independent reconstruction models are rarely benchmarked against Ground Truth (GT) data, with the notable exception of~\cite{williamsAssessingAccuracyAncestral2006}, where data is simulated along a phylogenetic tree using a structure-based and epistatic evolution model. 
In most other works, accuracy of the reconstruction is often solely measured according to the site-independent probability assigned to a given ancestral state~\cite{chantreauAsymmetricalDiversificationReceptorligand2019}. 
This raises the essential issue of the reliability of ASR methods. 
If site-independent reconstructions are only evaluated on their own probability distributions, what guarantee does one have to reconstruct the true ancestral sequence? 

Ideally, one would compare reconstructed ancestral sequences to known ancestors. Unfortunately,  no  complete ancestral DNA or protein sequences of sufficiently distant times are experimentally available; the oldest ancestral DNA sequence recovered and rebuilt from fragments, therefore containing little continuous information, dates only from a few million years ago~\cite{dalenDeeptimePaleogenomicsLimits2023}, 
which is nowhere near the timescale of the latest common ancestor of all life on Earth, estimated to have lived around 3.5 to 4 billion years ago~\cite{weissLastUniversalCommon2018}. Furthermore, known ancestral sequences tend to lie at modest Hamming distance from extant homologs relative to the full intra-family sequence variability, making the reconstruction problem comparatively trivial.

As a result, the use of forward simulators emerges as the only solution for efficient benchmarking, generating \textit{in silico} data at both the root and all internal nodes of a phylogenetic tree.
Many forward simulators over phylogenetic trees exist~\cite{fletcher2009indelible, ly2022alisim}, based on a wide class of substitution models~\cite{Dayhoff1978Model, whelanGeneralEmpiricalModel2001a, jonesRapidGenerationMutation1992}.
However, they usually suffer from two main limitations: (i)~ignoring the interaction between sites, and (ii)~lack of specificity for the gene or protein family being simulated. 
They are typically inferred on large datasets that span whole genomes and several hundred species, making them hardly fitting for such a precise task as functional sequence evolution. 
Unsurprisingly, they produce artificial sequences that are easily distinguishable from natural ones~\cite{trostSimulationsSequenceEvolution2024}.

A major shortcoming underlying both these well-known simulators and state-of-the-art reconstruction algorithms is the absence of explicit coevolutionary modeling~\cite{de2014empirical,starr2016epistasis,johnson2023epistasis}. 
Indeed, extant proteins are governed by stringent coevolutionary constraints, where interactions between residues fundamentally shape both the accessible sequence space and the resulting functional viability. 
As a result, the fitness effect of a mutation at a specific position is inherently conditional on the residues occupying other sites throughout the sequence. 
Consequently, this intricate network of coevolutionary interactions has been extensively explored through both experimental and computational lenses~\cite{domingo2019causes, chen2023understanding, cocco2018inverse}. 
Unfortunately, directly incorporating coevolution into reconstruction is computationally intractable due to the combinatorial complexity of site-site interactions. 

Several lines of research have emerged in response to this limitation. 
In particular, recent attempts have been made to leverage neural networks to encompass coevolution complexity. 
References~\cite{Ding2019deciphering, Hie2022evolutionary,Gorstein2025ancestral} notably propose to perform ASR in latent spaces inferred from sequence data. 
However, despite their increased complexity, these models still yield worse results than the classical site-independent method. Another recent work leverages transformer architectures~\cite{vaswani2017attention} to learn the transition probability between two sequences over a certain mutational time directly from data~\cite{Koehl2026deep}. 
Although these frameworks provide novel perspectives, they are not easily interpretable, as they bypass the specific biophysical rules governing sequence evolution. 
Moreover, they rely on expensive deep foundation models that require significant computational resources.
Simpler auto-regressive coevolutionary models can be used for ASR~\cite{pierre_arDCA}. 
While this approach produces interesting results at a reduced computational cost, the auto-regressive framework has the disadvantage of being non-Markovian and irreversible, making it biologically unrealistic.

This work proposes a different, bio-physically interpretable approach to overcome the site-independence assumption. 
The main achievement of this paper is the introduction of coevolutionary signal and epistasis into ASR in a computationally tractable manner. 
Indeed, as stated above, exact inference of ancestral sequences using coevolution models on large phylogenetic trees would require summing over an exponentially large sequence space, which is computationally prohibitive. 
Our solution is to obtain the single-site marginals of the posterior distribution using standard ASR, and to then use a coevolution-informed prior that epistatically couples the amino acids across sites, thus untangling the trade-off between tractable inference and realistic modeling of coevolution.
Importantly, our approach allows for biologically plausible sampling from the inferred ancestral probability distribution. 
Sampling a set of ancestral sequences, as opposed to focusing on the single MAP candidate, better represents the stochasticity of evolution and the potential diversity of properties of the ``true'' ancestral sequence~\cite{williamsAssessingAccuracyAncestral2006}. Importantly, the ability to sample from a realistic ancestral distribution gives the user a much better idea about ancestral uncertainty and potential diversity at the root.

To verify the accuracy of our coevolution-informed ASR procedure, we introduce a new benchmarking method based on a  Direct Coupling Analysis (DCA)~\cite{cocco2018inverse} forward simulator~\cite{DiBari2026modeling}. 
This serves as a more realistic and biologically grounded simulator with respect to site-independent traditional substitution models~\cite{ly2022alisim, fletcher2009indelible}.
In the context of DCA, functional epistatic constraints are represented by a coevolutionary model~\cite{DiBari2024, bisardi2022modeling} learned from protein family alignments.
Evolutionary trajectories produced by the DCA model  quantitatively match experimental evolution on short timescales~\cite{bisardi2022modeling, DiBari2024, DiBari2026modeling} while reproducing statistical properties of natural sequences and protein evolution on long time scales~\cite{Rossi_2025, de2020epistatic, alvarez2022novel}. 
In contrast to deep learning models, parameters of the DCA model are easily interpretable, with strong inferred pairwise interactions corresponding to physical contacts within the protein fold~\cite{weigt2009identification}. 
Most importantly, sequences sampled from the DCA model have been experimentally found to be functional~\cite{russ2020evolution, alvarez2024vivo, Lambert2024expanding, netti2026expanding}. 

Using the pipeline described above, we benchmark our coevolution-aware ASR against the basic site-independent Yang algorithm~\cite{yangNewMethodInference1995}, rather than more complex state-of-the-art pipelines such as IQ-TREE~\cite{IqtreeIqtree32025} with branch length resampling or mixed substitution models. This choice is deliberate: we seek a simple, explicit, fast baseline on top of which we can isolate the effect of adding coevolutionary constraints to a non-epistatic model, without confounding heuristics such as branch length rescaling or insertion-deletion events (replaced here by gaps in fixed-length sequences). We note, however, that our reshuffling procedure is not tied to any particular upstream method -- it takes as input any alignment of candidate ancestors sampled from a site-independent posterior, and can thus be applied downstream of any ASR method.

More specifically, we test all ASRs on a realistic phylogenetic structure inferred from extant protein data from two diverse protein families ($\beta$-lactamase and  DNA Binding Domain, see SI), playing on the scale of the tree, or average root-to-leaf distance, to simulate different evolutionary regimes. 
Schematically, a small evolutionary scale would lead to leaf sequences being clustered together - and thus to a small effective tree size; on the other hand, large evolutionary scales would produce very distant leaves, with little to no phylogenetic correlations, emulating the extreme case of a tree with no internal nodes, where the most recent common ancestor among any two leaves is the root. 
The idea here is to set up and present a principled and rigorous framework for the evaluation of ASR; we focus on a single, realistic tree structure, and dilate or compress evolutionary scales, from different starting root sequences, to explore a wide array of typical ASR regimes. 

Our results focus on the $\beta$-lactamase protein family (and extend in SI to the DNA Binding Domain) and demonstrate that enforcing coevolutionary interactions between amino acids at reconstruction time substantially improves ancestral sequence reconstruction when the true ancestor has strong epistatic constraints that limit its mutability as compared to the entire family.
Finally, we propose alternative quantitative readouts to mitigate biases inherent to site-independent ASR, and to better capture biological relevance. 
Altogether, this work highlights the importance of explicitly modeling coevolutionary interactions in ASR, and provides a principled method for reconstructing ancestral sequences that are both evolutionarily coherent and statistically comparable with natural sequences. 

\begin{figure*}
  \centering
\includegraphics[width=0.95\textwidth]{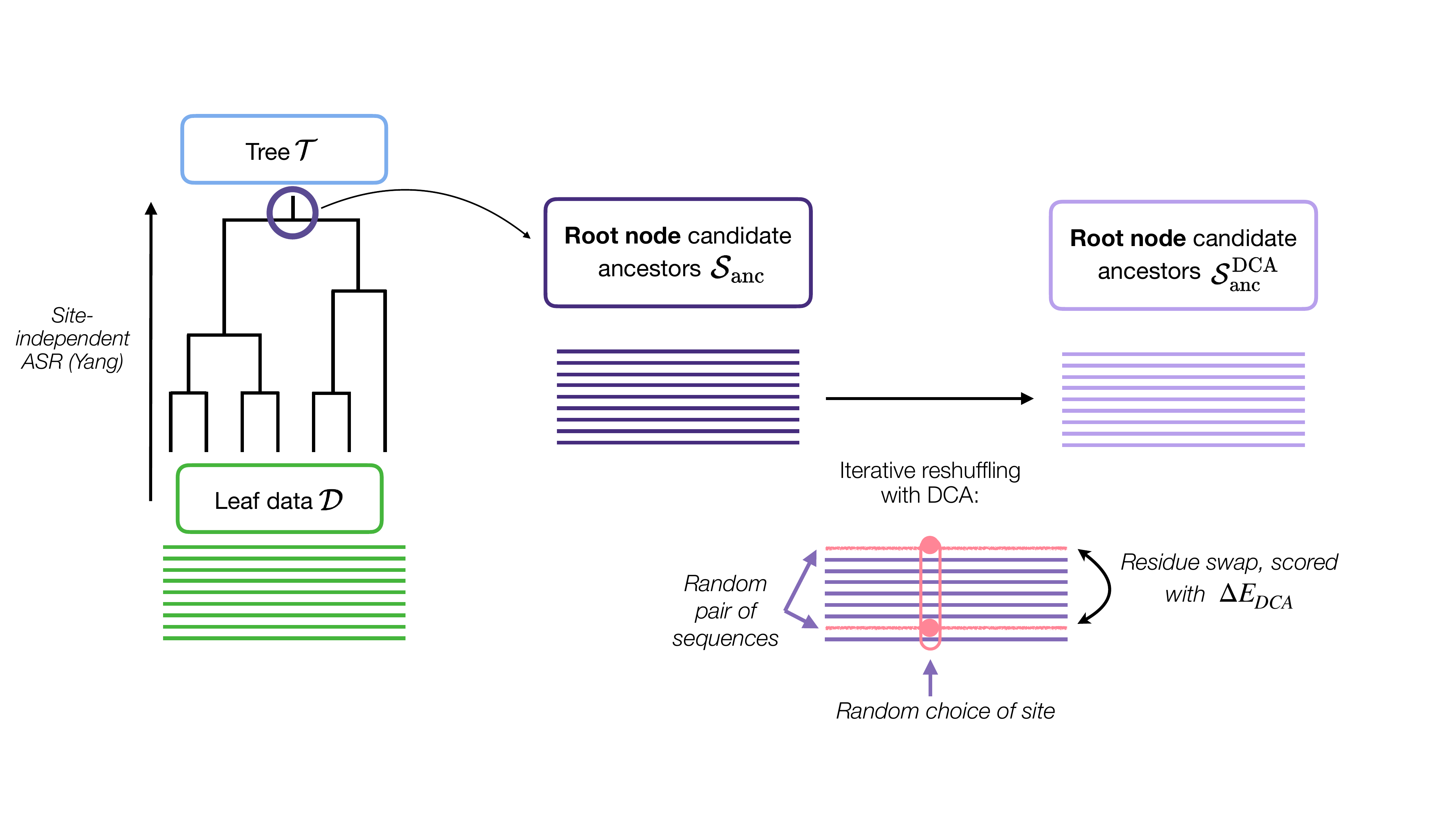}
  \caption{
      \textbf{Co-evolution aware ASR algorithm.} 
      The procedure reconstructs ancestral states through three main steps: (i) Site-independent reconstruction, where per-site posterior amino acid distributions are inferred at the root node using a profile model; (ii) Sampling of candidate ancestors, where a set of candidate root sequences is generated to explore the ancestral sequence space; and (iii) Coevolutionary reshuffling, where candidate sequences are iteratively optimized via residue swaps to reconcile phylogenetic history with the coevolutionary interactions encoded in the protein family.   
  }
  \label{fig:coevolution_asr_scheme}
\end{figure*}

\section{Results}
\subsection{Coevolution-aware ASR algorithm}\label{subsec:coevolution_aware_ASR_algo}
The central contribution of this work is a coevolution-aware ASR method that explicitly accounts for site-site interactions by integrating DCA~\cite{cocco2018inverse} constraints into a site-independent posterior distribution, enabling the sampling of biologically realistic ancestral sequences that reflect the structural and functional constraints of natural proteins -- a capability absent from standard site-independent approaches. The algorithm proceeds in two steps: (i) standard Bayesian site-independent ASR, followed by (ii) a reshuffling procedure that imposes coevolutionary constraints on the reconstructed sequences. Although applicable to any internal node of the phylogeny, we focus on root reconstruction throughout.  

As mentioned above, we chose to rely on DCA~\cite{cocco2018inverse} to incorporate coevolutionary signals into ASR. DCA models the statistical distribution of functional protein sequences via a probability distribution:
\begin{equation}
    \DCAprob(\vect{s}) \propto \exp\left[ - E_\text{DCA}(\vect{s}) \right],
\end{equation}
where
\begin{equation}
E_\text{DCA}(\vect{s})
= - \sum_{i=1}^L h_i(s_i)
  - \sum_{i < j} J_{ij}(s_i, s_j).
\end{equation}
The energy $E_\text{DCA}$ involves both site-specific ($h$) and pairwise coupling ($J$) parameters. Importantly, low-energy sequences are more likely to be phenotypically functional in experimental validation~\cite{russ2020evolution, alvarez2024vivo, Calvanese2025generating, Lambert2024expanding}. 

Unfortunately, since the DCA energy function couples sites across the sequence, marginalizing over all possible sequence states is computationally intractable ($\sim q^L$ states), precluding the direct use of DCA as a transition model over phylogenetic trees.

To circumvent this limitation, our coevolution-aware ASR procedure follows three steps (Fig.~\ref{fig:coevolution_asr_scheme}):
\begin{enumerate}
    \item Site-independent reconstruction (Section Methods~\ref{method:site_independent_ASR}). 
    We first perform standard site-independent ASR using the Yang et al. algorithm~\cite{yangNewMethodInference1995} to infer the per-site posterior distribution of amino acids at the root node. Rather than using a traditional substitution model (WAG~\cite{whelanGeneralEmpiricalModel2001a}, LG~\cite{leGascuelModelingProteinEvolution2012}, or JTT~\cite{jukesEvolutionProteinMolecules1969}), or even more complex ones that incorporate heuristics such as branch length resampling, insertions and deletions, etc., we use a profile model in which the stationary distribution at each site is set to the observed per-site amino acid frequencies in the leaf alignment $\Data$. This procedure is simple, basic and incredibly fast when parallelized over sites, and yields per-site posterior distributions $\postpersite(\cdot \given \DataVec{i}, \Tree, \vect{\theta_i})$ at the root node. 

    \item Sampling of candidate ancestors. We draw $M = 1000$ candidate root sequences from the site-independent posterior $\Post(\cdot \given \Data, \Tree, \vect{\theta})$, obtaining the candidate set $\ancestralset$.
    \item Coevolutionary reshuffling (Section Methods~\ref{method: co_evolution_aware_ASR}). A site $i$ and two sequences $\Seq^\alpha$, $\Seq^\beta \in \ancestralset$ are chosen at random, and their amino acids at site $i$ are swapped to obtain sequences $\hat{\Seq}^\alpha$ and $\hat{\Seq}^\beta$. The swap is accepted with probability $\min\bigl(1, e^{-\Delta E_\text{DCA}^{i,\alpha\beta}/T}\bigr)$, where 
    \begin{equation}
    \begin{split}
        \Delta E_\text{DCA}^{i,\alpha\beta} =
        & \bigl[ E_\text{DCA}(\hat{\Seq}^\alpha) + E_\text{DCA}(\hat{\Seq}^\beta) \bigr] \\
        & - \bigl[ E_\text{DCA}(\Seq^\alpha) + E_\text{DCA}(\Seq^\beta) \bigr],
    \end{split}
    \end{equation}
    and $T$ is an effective temperature controlling selection stringency. This is repeated for many cycles, producing the final set of ancestral candidates $\DCAset$.
\end{enumerate}

Because swaps occur only between amino acids at the same site, the per-site amino acid frequencies inferred in step~1 are exactly preserved throughout reshuffling. Consequently, $\DCAset$ satisfies the site-independent posterior $\postpersite$ while minimizing the DCA energy -- that is, maximizing the likelihood of the alignment under the DCA model. The result is a set of candidate ancestors that is both phylogenetically consistent and biologically plausible, combining the computational efficiency of the Yang dynamic programming algorithm with coevolutionary constraints learned by DCA from $\Data_\text{extant}$. More details are given in Methods (Section~\ref{method: co_evolution_aware_ASR}); the definitions of the variables are collected in Table~\ref{table:variables} listed later. 

Notably, any Bayesian ASR method can be applied to obtain the site-independent posterior distribution; our coevolution-aware ASR procedure only needs as input an alignment of ancestral candidates sampled from the ancestral posterior. The process of obtaining this posterior does not affect our ability to introduce coevolutionary couplings in ancestral candidates. 

\subsection{A realistic evolver to benchmark ASR}

To rigorously assess the performance of our coevolution-aware reconstruction method, we require a controlled setting in which the true ancestral sequences are known. We therefore developed a dedicated benchmarking framework designed to provide a biologically meaningful context for ASR evaluation, with particular emphasis on realistic evolutionary simulations.

Because the DCA model is generative of phenotypically functional sequences and reproduces both site-wise and pairwise amino acid statistics of the extant family, repeatedly simulating evolution from the same root sequence on independent lineages yields a leaf distribution that matches the statistical ensemble of natural sequences. 
Simulating along the branches of a phylogenetic tree rather than over indipendent lineages adds an extra layer of realism: the finite branch lengths introduce the inter-sequence phylogenetic correlations characteristic of real biological sequence data, while the leaf sequences remain functional and representative of the protein family with high probability. Consequently, sequences simulated across the entire phylogeny preserve the statistical properties and structural constraints observed in natural proteins, while explicitly incorporating coevolutionary patterns and phylogenetic relations.

The evolver used in this study follows a framework developed in previous work~\cite{de2020epistatic, alvarez2024vivo, DiBari2024, Rossi_2025, DiBari2026modeling} to simulate sequence evolution under coevolutionary constraints. Starting from the ground truth root sequence $\GT$, descendant sequences are evolved recursively along the rooted phylogenetic tree via a discrete-time Markov chain. At each step, a single amino acid mutation is proposed at a random site $i$ and the move is accepted depending on the DCA energy change induced by the mutation, which in turn depends on the current amino acid states at all other sites $j \neq i$ through the pairwise couplings $J_{ij}$ (see Eq.~\ref{eq:energy}, Methods Section~\ref{method:DCA_inference}). This process is iterated over each branch for a number of mutation attempts proportional to its length, naturally encoding the phylogenetic timescale into the evolutionary dynamics.

To explore how ASR performs across various evolutionary timescales, we apply a global scaling factor $\muDCA$ to the branch lengths of the phylogenetic tree, simulating evolutionary divergence ranging from nearly identical leaves to highly divergent ones while keeping the phylogenetic topology fixed. 
Concretely, the total number of mutation attempts over a branch of length $t$ is $\muDCA L t$, where $L$ is the sequence length. 
Branch lengths are normalized so that the average root-to-leaf distance $\langle t_{R,\text{leaf}}\rangle_\text{leaf} = 1$, meaning $\muDCA = 1$ corresponds to one proposed mutation per site along the average root-to-leaf path. 
More broadly, $\muDCA$ provides a controlled axis along which to probe different evolutionary regimes: low values correspond to shallow phylogenies with closely related sequences, while large values correspond to the deep divergence observed in extended protein families spanning Archaea, Bacteria, and Eukaryota.

On the $\beta$-lactamase family (see below), FastTree~\cite{priceMorgannpriceFasttree2025} yields $\langle t_{R,\text{leaf}}\rangle = 2.61$ before normalization, so that in a site-independent framework $\muDCA = 2.61$ reproduces the extant sequence distribution at the leaves. The introduction of epistatic couplings in the DCA model slows sequence divergence, requiring larger $\muDCA$ values to reach the same distribution; the correspondence between $\muDCA$ and the mutation rate of the Yang site-specific and site-independent propagator $\mu_\text{site-independent}$ is given in SI (Fig.~S4).

To benchmark our coevolution-aware algorithm in different evolutionary regimes, we choose five distinct root sequences, among the extant ones, with different levels of mutability.
More precisely, based on previous studies ~\cite{vigue2022deciphering, DiBari2024, Rossi_2025, pagnani2025generative}, we use the DCA model to compute a sequence-specific mutability metric, called Context-Dependent Entropy (CDE), that determines the speed of evolution starting from the sequence of interest (see Methods~\ref{method:choice_of_root_sequences} for details).
Indeed, highly mutable sequences will diverge quickly over time, following an evolutionary dynamic that is similar to that of a site-independent, site-specific evolver, whereas weakly mutable sequences, in which epistatic constraints limit the number of tolerated mutations, will experience a much slower evolution.

\subsection{Analyzing forward evolution on the phylogenetic tree}
\label{subsec:one}

\begin{figure*}
  \centering
\includegraphics[width=0.95\textwidth]{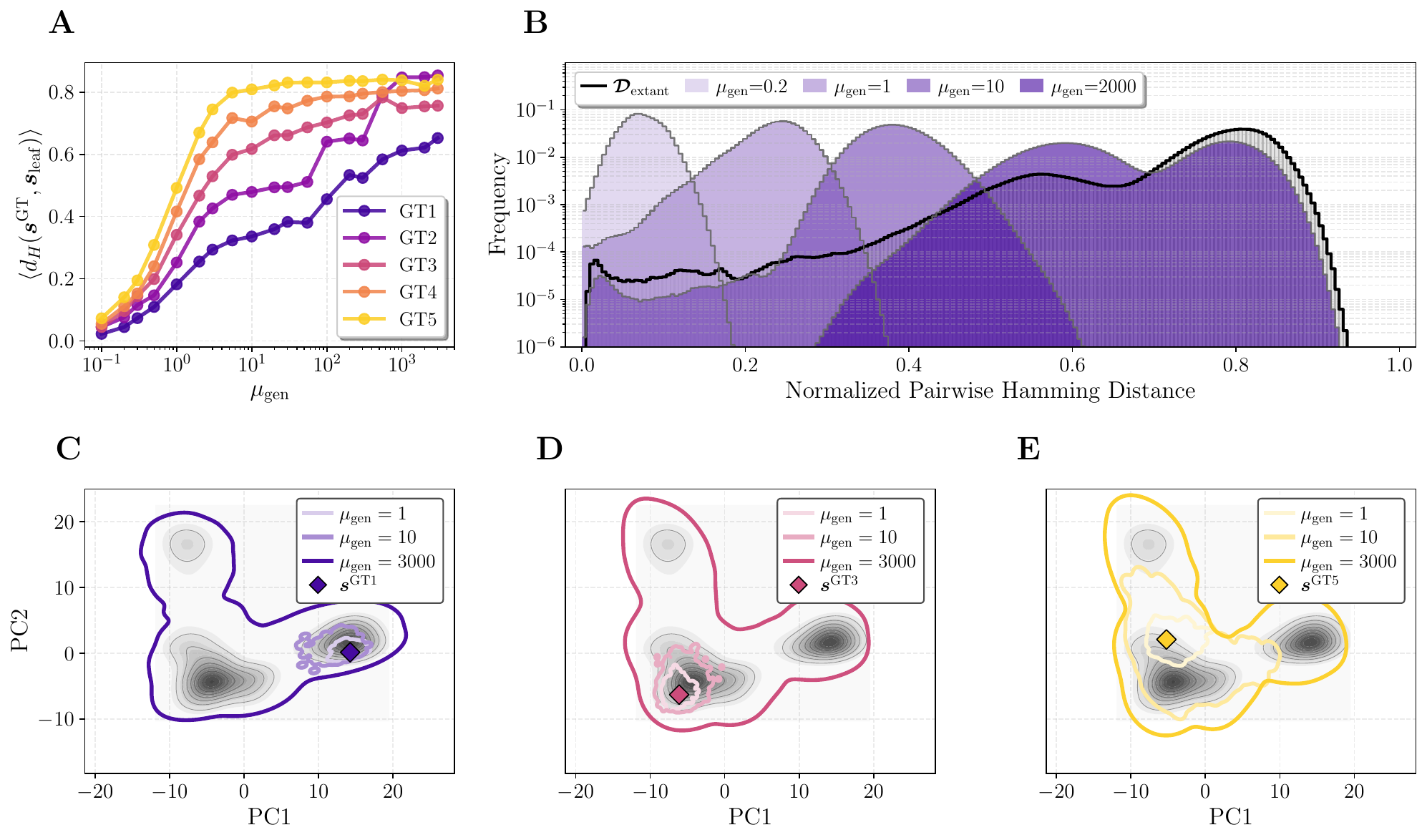}
  \caption{
      \textbf{Effect of sampling time on forward generated sequences.} 
      Sequences were simulated with the DCA model along the phylogenetic tree of the $\beta$-lactamase family, using increasing $\muDCA$ values. (A) Average leaf–root Hamming distance for simulated data as a function of $\muDCA$. (B) Distribution of pairwise Hamming distances $d_\text{H}(\bm{s}^i, \bm{s}^j), \text{ with } \bm{s}^i, \bm{s}^j \in \Data$ among simulated leaves (shown here for $\GTindexed{1}$). The gray histogram outlined in black represents the pairwise Hamming distance distribution of the extant sequences of the $\beta$-lactamase family. (C,D,E) Principal Component Analysis (PCA) of the sampled $\Data$ for $\GTindexed{1}, \GTindexed{3}, \GTindexed{5}$ respectively. Extant sequences $\Data_\text{extant}$ are plotted as the gray density distribution, on principal components 1 and 2. Colored outlines show the exploration of PCA space for different evolutionary scales $\muDCA$. 
  }
  \label{fig:forward_evolution}
\end{figure*}

We performed our analysis on the $\beta$-lactamase protein family (see SI for results on the DNA Binding Domain).
The alignment of natural sequences was used to infer a phylogenetic tree using FastTree~\cite{priceMorgannpriceFasttree2025} as well as a DCA and site-independent model for ASR,  ensuring consistency between the inferred evolutionary history and the statistical constraints of the protein family (see Methods).
Five root sequences were selected from the alignment using the DCA model and the procedure described above. 
Starting from each root sequence, we simulated sequence evolution along the tree $\Tree$ using the DCA-based evolver (Fig.~\ref{fig:forward_evolution}). 
Unlike site-independent traditional substitution models which tend to drift into non-functional regions of sequence space (see SI, Fig.~S6), the DCA-based evolver reproduces the empirical single-site and pairwise amino acid frequencies of extant sequences, enabling reliable exploration of the functional sequence space over long evolutionary timescales.

The five root sequences $\GTindexed{1}, \ldots, \GTindexed{5}$ were selected at increasing mutational tolerance (see Methods~\ref{method:choice_of_root_sequences}), with $\GTindexed{1}$ and $\GTindexed{5}$ being respectively the least and most mutable extant sequences in the alignment. 
Panel~A shows that the average normalized Hamming distance  $\langle d_{\text{H}}(\GT, \Seq_\text{leaf})\rangle$ increases monotonically with $\muDCA$ as expected. 
Note that all Hamming distances in this work are normalized by the length of the sequences $L$.
By construction, the evolver converges towards the sequence distribution of the training set, so that as $\muDCA \to \infty$ the distance saturates to the average distance from each extant sequence to the chosen root.
Both the convergence rate and the limiting Hamming distance vary with the chosen root~$\GT$: for instance, saturation is considerably slower for $\GTindexed{1}$ than for other roots~\cite{Rossi_2025}.
This dependence on root mutability is further reflected in panels~\ref{fig:forward_evolution}C--E, which represent sequence space exploration using the first two principal components of the alignment. 
Exploration of this space by the evolver differs across roots both in regions explored and in diversification speed.

Panel~\ref{fig:forward_evolution}B focuses on the distribution of pairwise Hamming distances among simulated leaves (shown here for $\GTindexed{1}$). Notably, no single value of $\muDCA$ exactly reproduces the empirical pairwise distance distribution observed in natural sequences. At low $\muDCA$, sequences remain similar, resulting in a narrow, low-distance pairwise distribution. Intermediate values of $\muDCA$ better capture the left tail of the empirical distribution, corresponding to closely related sequences and thus reflecting phylogenetic relationships. In contrast, high $\muDCA$ values better reproduce overall diversity - as expected from the DCA model, which is calibrated to match the overall statistics of natural sequence variability - but approach sequence independence, effectively erasing phylogenetic correlations. 

This mismatch arises from the assumptions underlying the training of the DCA model, which treats natural sequences as independent samples from an unknown probability distribution, thus neglecting phylogenetic correlations~\cite{cocco2018inverse}; previous work has shown that this approximation has a limited impact on the quality of coevolutionary inference~\cite{hortaEffectPhylogeneticCorrelations2021}. During training, sequence reweighting is used to partially account for phylogenetic relationships and reduce bias, by down-weighting closely related sequences corresponding to the left tail in panel~\ref{fig:forward_evolution}B. When sequences are subsequently evolved along an explicit phylogeny, the phylogenetic signal is transiently recovered by reconstructions, but at long times the pairwise distance converges towards that of an independent-lineage DCA sample, where the left tail is suppressed.

\subsection{Effect of root mutability and divergence time on ASR}

Having established our benchmarking framework, we can now focus on assessing the accuracy of site-independent ASR methods. 
More specifically, we are interested in heuristically narrowing down the evolutionary regimes (governed by the evolutionary scale $\muDCA$, and influenced by root mutability) in which ASR is possible and non-trivial. 
While previous work has established the exact conditions for recoverability in the case of simplified binary models on trees~\cite{evansBroadcastingTreesIsing2000}, extending these results to multi-state alphabets and the heterogeneous constraints of protein sequence space remains analytically inaccessible, leaving us without comprehensive theoretical guidance. 

To answer this question, we use step 1 of the procedure referenced in Section~\ref{subsec:coevolution_aware_ASR_algo} which yields the site-independent ancestral posterior distribution $\Post(\cdot \given \Data, \Tree, \vect{\theta})$, subsequently referenced as $\Post$ for brevity. From $\Post$, we then derive the Maximum A Posteriori sequence $\MAP$ (see Equation~\ref{eq:MAP} or Table~\ref{table:variables}), which assigns to each site the state with the largest site-independent posterior probability.

The consensus $\Cons$ of the leaves $\Data$, which we use as a baseline for comparison with the MAP reconstruction, is defined at each site $i$ as:
\begin{equation} \label{eq:consensus}
        s^\text{cons}_i
        =
        \arg\max_{a \in \mathcal{A}}
        f_i(a),
    \end{equation}
where $f_i(a)$ is the frequency of amino acid $a$ at site $i$ in the sequences $\Data$. This sequence amounts to taking the most frequent state at each site. 
In contrast to MAP, the consensus can be seen as the most likely candidate ancestor in the absence of phylogeny and of site-specific biases; that is, the sequence that maximizes the site-independent posterior probability if one considers the leaves to be the result of evolution on independent lineages. 
Comparing these MAP and consensus reconstructions allows us to quantify the additional information provided by phylogeny beyond simple frequency aggregation.
All definitions of the relevant sequences can be found in Table~\ref{table:variables} below. 

\begin{table}[H]
\centering
\small
\caption{Definition of ASR sequences and candidate ensembles.}
\label{table:variables}
\begin{tabular}{p{0.18\linewidth} p{0.70\linewidth}}
\hline
\textbf{Symbol} & \textbf{Definition} \\
\hline
$\GT$
& Ground-truth ancestral sequence used as the root of forward simulations, drawn from $\Data_\text{extant}$. \\
$\MAP$
& $\displaystyle \arg\max_{\vect{s}} \Post(\vect{s} \given \Data, \Tree, \vect{\theta})$, maximizer of the site-independent posterior at the root. \\
$\Cons$
& $\displaystyle s^\text{cons}_i = \arg\max_{a \in \mathcal{A}} f_i(a)$, consensus sequence obtained from $\Data$ by selecting the most frequent amino acid at each site. \\
\hline
$\ancestralset$
& $\displaystyle \{\vect{s}^{(k)}\}_{k=1}^M \sim \Post(\cdot \given \Data, \Tree, \vect{\theta})$, with $M=1000$, sampled from the site-independent posterior. \\
$\DCAset$
& $\displaystyle \{\vect{s}^{(k)}\}_{k=1}^M$, obtained by reshuffling $\ancestralset$ to impose coevolutionary constraints via a DCA model inferred from $\Data_\text{extant}$. \\
\hline
\end{tabular}
\end{table}

\begin{figure*}[t]
  \centering
  \includegraphics[width=0.95\textwidth]{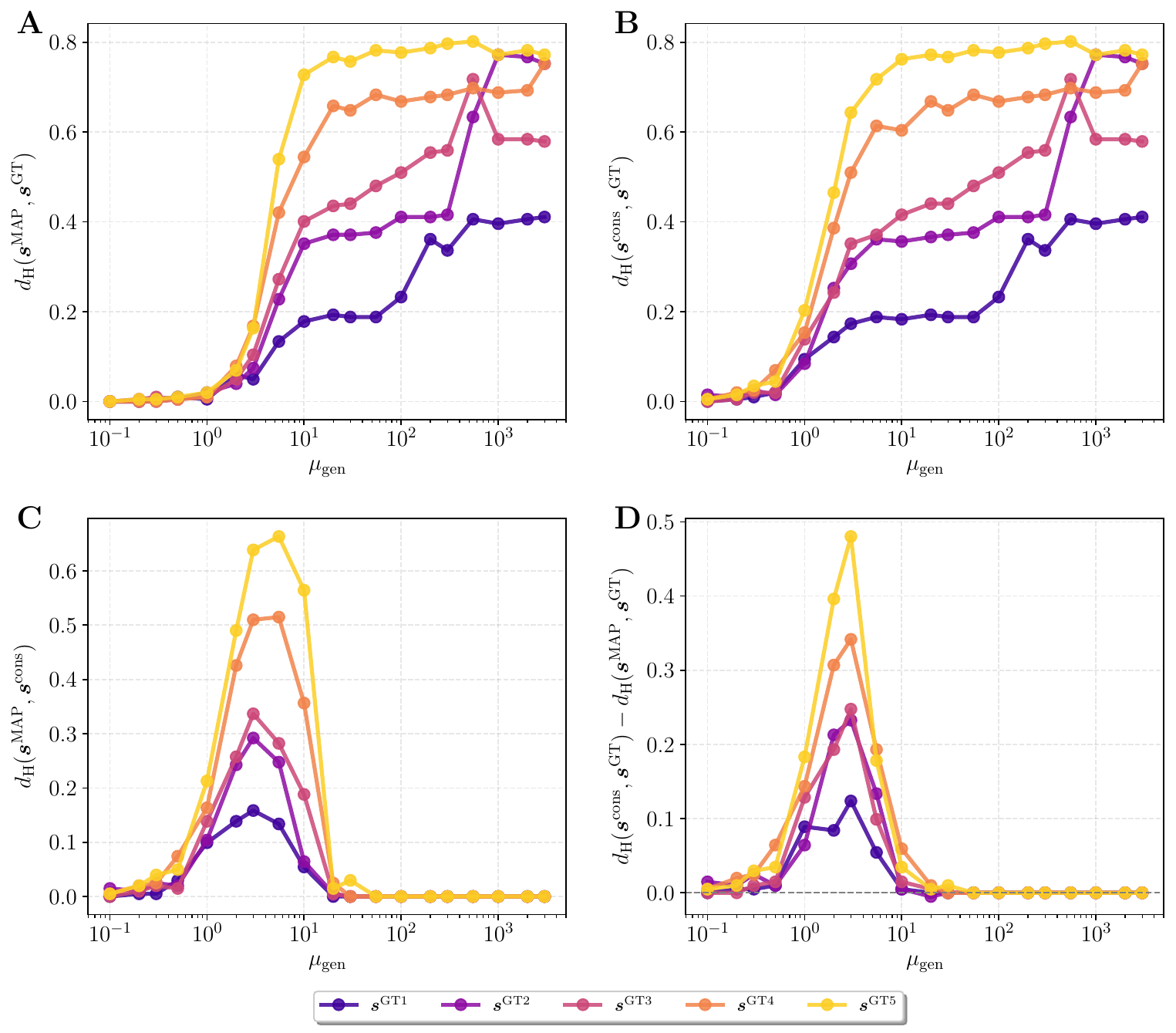}
  \caption{
      \textbf{Effect of evolutionary time $\muDCA$ and root mutability on ASR accuracy in the $\beta$-lactamase family.}
      Normalized Hamming distance plots between Maximum a Posteriori reconstructed ancestor $\MAP$, GT ancestor $\GT$, and consensus sequence of the leaves $\Cons$ as a function of evolutionary time $\muDCA$.
      (A)~Hamming distance between MAP and GT. 
      (B)~Hamming distance between consensus sequence from the leaves and GT. 
      (C)~Hamming distance between MAP reconstruction and the consensus from the leaves. 
      (D)~Difference between the Hamming distances $d_{H}(\Cons,\GT)-d_{H}(\MAP,\GT)$. Positive values mean MAP reconstruction is closer to GT than the leaf-consensus.
  }
  \label{fig:mu_vs_asr}
\end{figure*}

Panel~\ref{fig:mu_vs_asr}A shows the normalized Hamming distance $d_\text{H}(\MAP, \GT)$ between MAP and GT sequences as a function of $\muDCA$. 
As expected, reconstruction accuracy decreases with increasing $\muDCA$ as the root becomes more evolutionarily distant from the leaves.
Interestingly, the accuracy strongly depends on ancestral sequence mutability. 
For highly mutable roots (GT4 and GT5), the divergence between MAP and GT increases rapidly with $\muDCA$, whereas weakly mutable roots (GT1 and GT2) exhibit slower degradation in performance. The MAP–GT distance is nearly monotonic with respect to the mutability of the original root sequence, indicating that stronger epistatic constraints enhance ancestral recoverability~\cite{Rossi_2025}, with less mutable roots being more accurately reconstructed. 

Similarly, panel~\ref{fig:mu_vs_asr}B displays the Hamming distance between the leaf consensus $\Cons$ and the GT ancestor $\GT$. 
The same trend is observed: high mutability leads to faster divergence from the true root.

Panel~\ref{fig:mu_vs_asr}C address directly the importance of phylogeny by showing the distance between $\MAP$ and $\Cons$. 
In the interval $\muDCA \in \sim[1, 100]$, $\MAP$ and $\Cons$ clearly differ, demonstrating that phylogenetic structure modifies the inferred ancestor. 
Notably, the MAP–consensus discrepancy is largest for mutable roots ($\GTindexed{4}$ and $\GTindexed{5}$), precisely where coevolutionary effects are weaker. 
In such regimes, phylogenetic correlations carry essential information not captured by simple residue frequencies.
Conversely, for weakly mutable roots ($\GTindexed{1}$ and $\GTindexed{2}$), coevolutionary constraints keep leaf sequences in a restricted region of sequence space, where the consensus $\Cons$ already lies close to the true ancestor and the $\MAP$ reconstruction offers relatively little improvement in Hamming distance accuracy.

Finally, Panel~\ref{fig:mu_vs_asr}D confirms that the MAP–consensus differences observed in Panel~\ref{fig:mu_vs_asr}C correspond to genuine improvements in reconstruction accuracy. 
The quantity 
\begin{equation}
d_\text{H}(\Cons,\GT) - d_\text{H}(\MAP,\GT)
\end{equation}
is largely positive throughout the relevant range of $\muDCA$ values, indicating that MAP reconstruction is systematically closer to the true ancestor $\GT$ than the consensus $\Cons$. 
Thus, phylogenetic inference does not merely produce alternative reconstructions, but improves proximity to the ground truth.
While curves in Panels~\ref{fig:mu_vs_asr}A and~\ref{fig:mu_vs_asr}B can look very similar, the MAP reconstruction retrieves information that the consensus sequence simply does not. 
Indeed, looking at proximity between MAP and GT (see SI Fig.~S5) and consensus and GT, as a function of divergence between root and leaves, one sees clearly that $d_\text{H}(\Cons,\GT)$ is simply a linear function of root-leaf divergence $\langle d_\text{H}(\GT, \vect{s}_\text{leaf})\rangle$, regardless of root mutability.
On the opposite, the MAP reconstructed ancestor shows better proximity to $\GT$, and differing results as a function of $\GT$ mutability. 

This analysis allows us to identify three regimes. At small evolutionary scales ($\muDCA \lesssim 1$), leaves remain so close to the root that phylogenetic inference becomes unnecessary: both $\MAP$ and $\Cons$ approximate the ancestor equally well. 
Second, when $\muDCA \gtrsim 100$ and the evolutionary scale is very large, sequences have diverged so much from the ancestor that neither $\MAP$ nor $\Cons$ can reliably recover the root $\GT$.  
The intermediate regime $\muDCA \in [1, 100]$  emerges as the regime where phylogenetic inference provides non-trivial information beyond the leaf consensus.
Based on this analysis, we use it as the relevant interval in the remainder of the manuscript. 

Importantly, this analysis also reveals how mutability of the ancestral sequence modulates ASR accuracy.  For highly-mutable sequences (weakly constrained by coevolutionary interactions) the reconstruction quality is worse at fixed $\muDCA$ (Panels~\ref{fig:mu_vs_asr}A,B) and the MAP estimate is far more distant from the consensus with respect to what happens for mutable sequences (Panel~\ref{fig:mu_vs_asr}C). 
In addition, we can see that phylogenetic information significantly improves performance for mutable sequences (Panel~\ref{fig:mu_vs_asr}D).

On the other hand, one could attribute this dependency on root mutability as a simple scaling problem depending on the starting point. To dispel this hypothesis, we plotted in SI (Fig.~S5) the Hamming distance between $\MAP$ and $\GT$ as panel A, and the Hamming distance between $\Cons$ and $\GT$ as panel B, this time as a function of average root-to-leaf Hamming distance. And while the distance between $\Cons$ and $\GT$ follows a linear function of root-to-leaf distance, and seems to be independent of the starting point, the distance between $\MAP$ and $\GT$ is very much root-dependent, and scales differently depending on the GT starting point. 

\subsection{Independent-site and coevolution-aware\\ ASR results}

Having restricted the evolutionary scale to ${\muDCA \in [1, 100]}$, we apply our coevolution-aware ASR procedure and benchmark it against both the GT ancestor $\GT$ and the Yang et al.\ site-independent ASR~\cite{yangNewMethodInference1995}.

\begin{figure*}[t]
  \centering
  \includegraphics[width=0.95\textwidth]{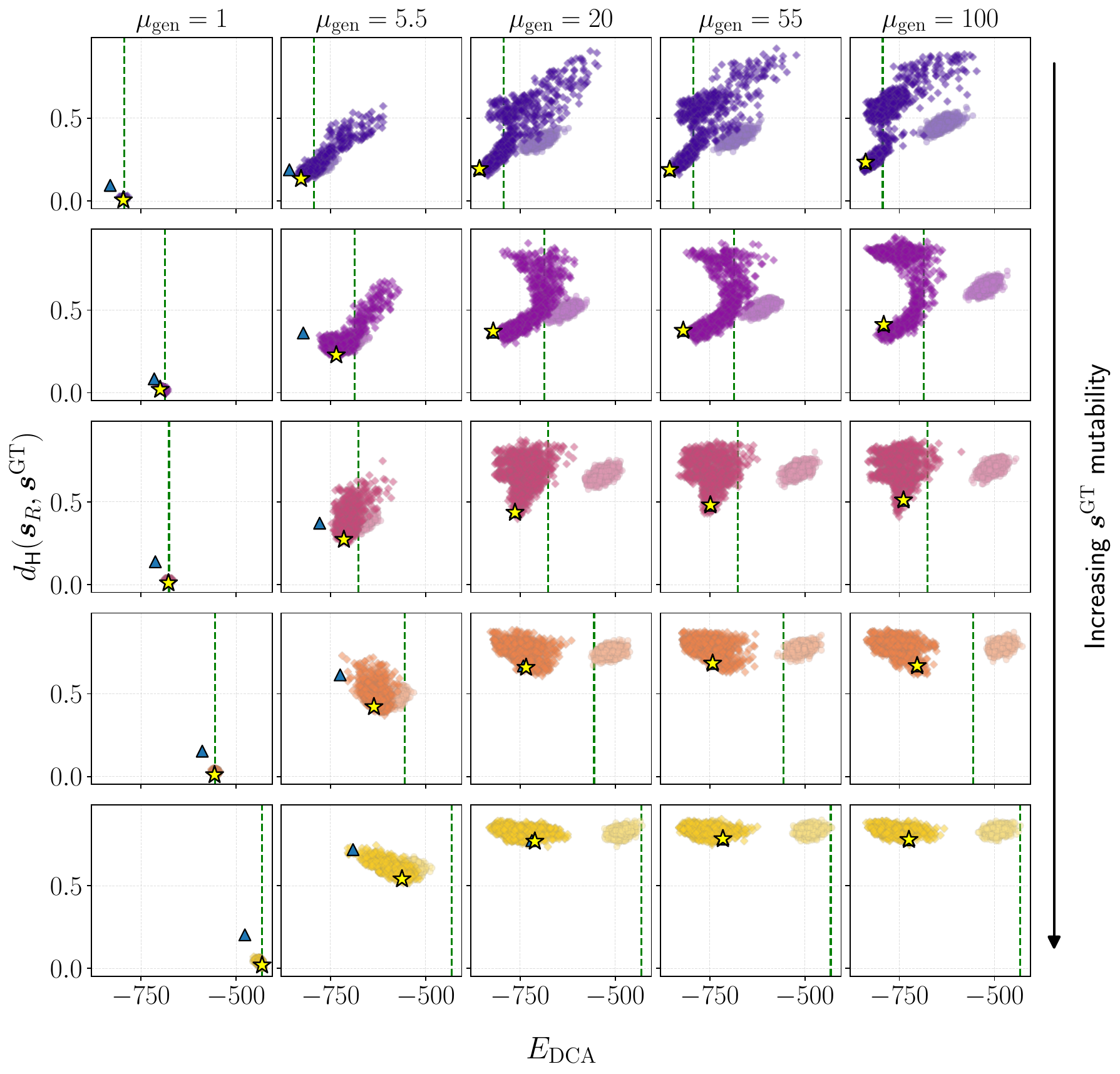}\\[1em]
  \includegraphics[width=0.95\textwidth]{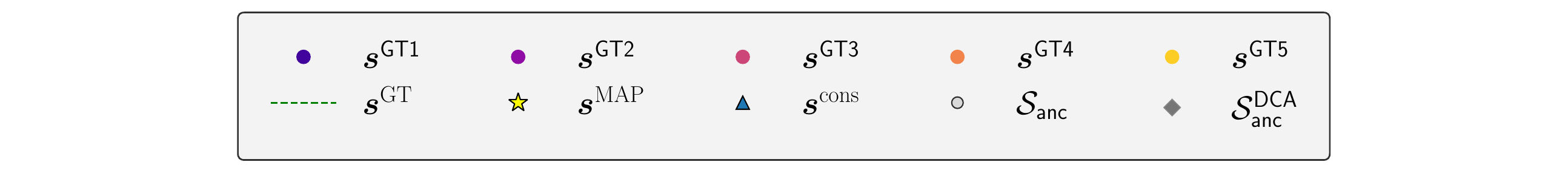}
  \caption{
      \textbf{Quantitative assessment of ancestral sequence reconstructions.}
      Comparison of $\GT$, $\MAP$, $\Cons$, ancestral samples $\ancestralset$ from the site-independent posterior $\Post$, and coevolution-aware ancestral samples $\DCAset$ across evolutionary times and GT roots.
      Sequences are evaluated using Hamming distance to $\GT$ (vertical axis) and DCA energy(horizontal axis). GT sequence $\GT$ is represented as the green vertical dotted line, MAP sequence $\MAP$ as the star, and consensus of the leaves $\Cons$ as the blue triangle. Two clouds of points are plotted: on each panel, the lighter cloud (individual shape: circle) is the candidate-ancestor set $\ancestralset$  sampled from the site-independent posterior $\Post$, while the cloud of darker shade (individual shape: diamond) represents the coevolution-aware candidate ancestors from the reshuffled set $\DCAset$.}
  \label{fig:Fig.4}
\end{figure*}

Within $\muDCA \in [1;100]$, we next evaluate ancestral sequence reconstruction at the root of the fixed $\beta$-lactamase phylogeny over: (i)~$\ancestralset$,
(ii)~$\DCAset$,
(iii)~$\MAP$, 
(iv)~$\Cons$,
against $\GT$, the GT sequence used as root. Again, all these quantities are summarized in Table~\ref{table:variables}.
In Fig.~\ref{fig:Fig.4} we plot each of these reconstructed sequences on two axes. 
The y-axis measures the Hamming distance $d_\text{H}(\RootSeq, \GT)$ from $\GT$, while the x-axis shows the DCA energy $E_\text{DCA}$.
The former gives us the proximity of reconstructed root $\RootSeq$ to the GT ancestor in terms of reconstruction accuracy; while the latter quantifies the compatibility of $\RootSeq$ with the inferred coevolution fitness landscape.
It has been shown~\cite{russ2020evolution, Lambert2024expanding, netti2026expanding} that the DCA energy is a reliable proxy for functionality: low-energy sequences are statistically enriched in phenotypically functional variants in experimental assays, while high-energy sequences present a much lower -- though non-zero -- probability of being functional. The DCA energy should therefore be understood as a probabilistic indicator rather than a binary functional classifier.

The range of DCA energies for extant sequences is shown in SI (Fig.~S2), alongside energies of sequences sampled from the site-independent profile model. Sequences sampled without coevolutionary constraints systematically yield high DCA energies, falling in a region the DCA model considers low-probability -- though with a partial overlap with extant sequences. 
Notably, $\GTindexed{5}$ was selected as the extant sequence with the smallest CDE according to DCA, making it atypical both in terms of DCA energy and, as discussed in Section~\ref{subsec:picking_good_DCA_ancestral_sequences}, for structural inference using ESMFold~\cite{Lin2023evolutionary} or AlphaFold~\cite{jumper2021highly}. As a result, site-independent ancestral candidates with a lower DCA energy than $\GTindexed{5}$ should not be interpreted as likely functional -- lower energy than an atypical extant sequence is a weak signal at best.

Even when starting from the least mutable root $\GTindexed{1}$, both $E_\text{DCA}(\MAP)$ and $E_\text{DCA}(\Cons)$ are systematically lower than $E_\text{DCA}(\GT)$. 
That is, $\MAP$ and $\Cons$ appear \emph{over-optimized} under the DCA model, relatively to the true ancestral sequence $\GT$. This effect is observed across all root sequences and evolutionary times.
This systematic drift toward atypically low DCA energies suggests that MAP and consensus reconstructions bias sequences toward configurations that maximize compatibility with the inferred single-site statistics, thereby indirectly over-optimizing DCA energy. 

DCA energy has been shown to be inversely correlated with thermal stability~\cite{morcos2014coevolutionary}, so that the resulting sequences, having low DCA energy, may therefore exhibit artificially enhanced stability or activity, a phenomenon frequently reported in experimental studies of reconstructed ancestors~\cite{zhaoAncestralSequenceReconstruction2026,supekarAncestralSequenceReconstruction2026,spenceAncestralSequenceReconstruction2021,prakineeAncestralSequenceReconstruction2024}.

In contrast, samples from the site-independent posterior $\Post$ are typically both distant from $\GT$ and high in DCA energy, a feature that is often associated with reduced structural stability or loss of function~\cite{russ2020evolution} as discussed above. 
Thus, although sampling from the posterior allows characterization of variability in root residues, it tends to explore regions of sequence space that are unlikely under the global coevolutionary model, often leading to non-functional proteins.

Coevolution-aware candidate ancestors $\DCAset$ exhibit markedly different behavior. 
Because the reshuffling procedure preserves single-site frequencies, the improvement in DCA energy arises exclusively from the introduction of coevolutionary correlations between sites. As a consequence, while some candidate sequences move closer to the true ancestor in Hamming distance, others may move further away -- the procedure broadens the sequence distribution in Hamming space in order to enforce pairwise constraints and maintain intra-sequence coherence.

For weakly mutable roots at intermediate evolutionary timescales, the higher-likelihood cluster of $\DCAset$ sequences simultaneously approaches the GT in both Hamming distance and DCA energy, outperforming their site-independent counterparts. For highly mutable roots, however, the GT itself exhibits higher DCA energy than the bulk of the extant family (see SI Fig.~S3), so that the reshuffling procedure -- which by design drives sequences toward energies typical of the extant family -- systematically proposes candidates of lower energy than the GT. This is a purposeful and inherent feature of our algorithm, and structural analysis (Section~\ref{subsec:picking_good_DCA_ancestral_sequences}) shows that the best $\DCAset$ candidates achieve closer structural proximity to the GT than either the MAP estimate or site-independent samples. 

Intermediate regimes ($\muDCA \lesssim 20$) best underline the difference between ASR on weakly mutable and highly mutable roots.
For weakly mutable roots ($\GTindexed{1}$ and $\GTindexed{2}$), ancestral reconstruction accuracy is consistently higher than for other GT sequences (rows 3 to 5), in agreement with Fig.~\ref{fig:mu_vs_asr}. 
Strong coevolution signal constrains evolutionary trajectories to a restricted region of sequence space, improving the recovery of the ancestral sequence.
For highly-mutable roots ($\GTindexed{4}$ and $\GTindexed{5}$), phylogenetic signal decays more rapidly.

Beyond $\muDCA \gtrsim 20$, all reconstruction methods converge to similar Hamming distances from the GT. 
In this regime $\MAP, \Cons, \DCAset$ all exhibit DCA energies substantially lower than that of the GT. 
Surely, this is expected in the case of $\DCAset$, as DCA energy is explicitly optimized under the coevolution-aware  procedure, modulated by the resampling temperature $T$ (see Methods~\ref{method: co_evolution_aware_ASR}). 
For MAP and consensus, however, this shift emerges indirectly and reflects model mis-specification rather than explicit optimization.

These results reveal a systematic tension between sequence accuracy (Hamming proximity to GT) and optimality under the inferred DCA model. 
Site-independent MAP reconstruction tends to generate sequences that are overly compatible with the learned statistical landscape, potentially exaggerating functional properties. 
Sampling from the posterior distribution preserves variability but fails to enforce coevolutionary coherence. 
The coevolution-aware reconstruction, by contrast, enforces epistatic constraints among sites while preserving site-independent marginal statistics, yielding reconstructions that can be simultaneously closer to the GT in sequence space and more consistent with the global DCA energy functionality proxy.
This naturally raises a final question: given access only to reconstructed sequences (and not to the GT ancestor), can one identify principled criteria to select sequences that are close to the true ancestor  both in Hamming distance and in DCA energy range, thereby outperforming the site-independent MAP estimator?

\begin{figure*}[t]
  \centering
  \includegraphics[width=0.95\textwidth]{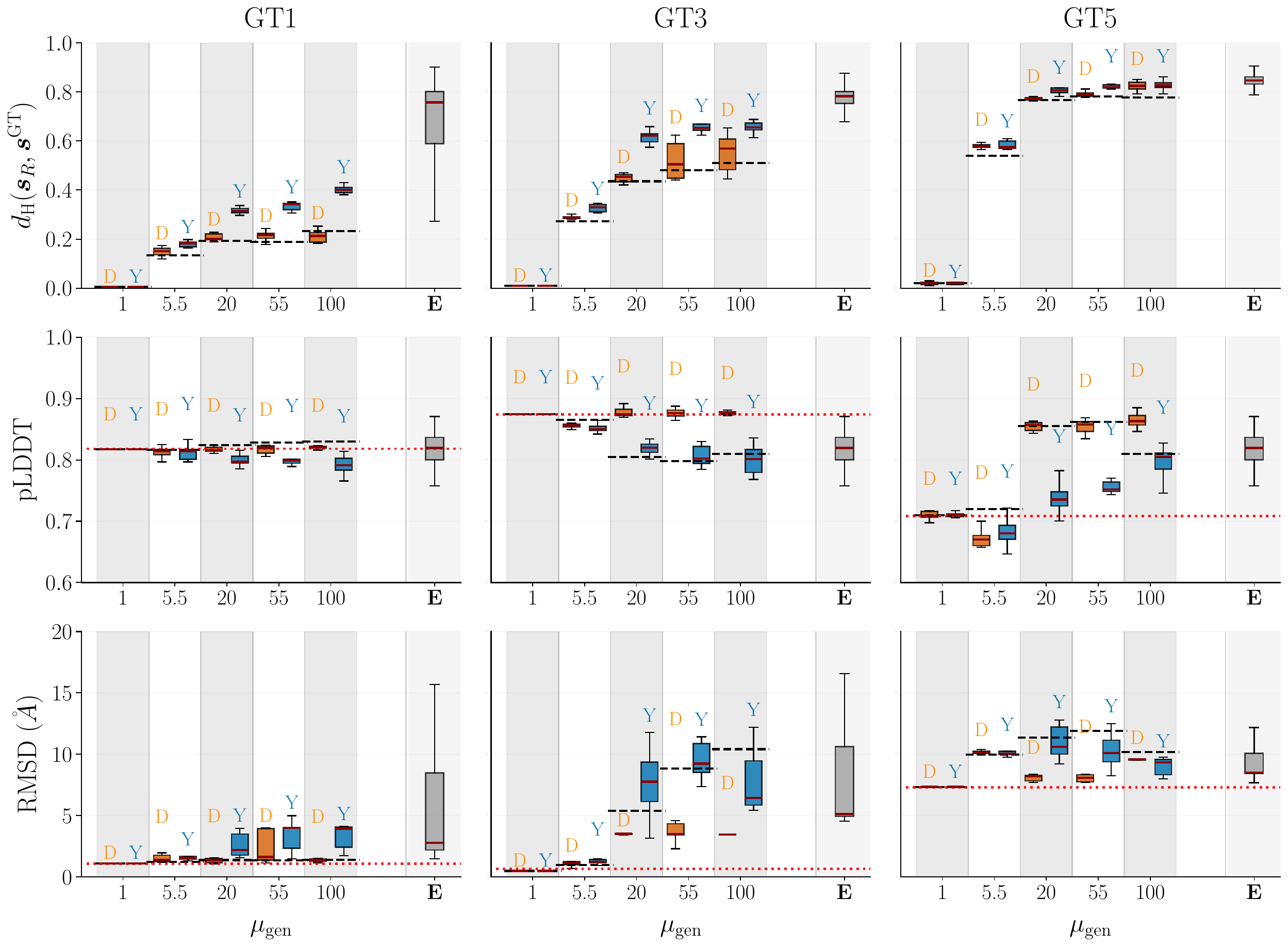}
  \includegraphics[width=0.95\textwidth]{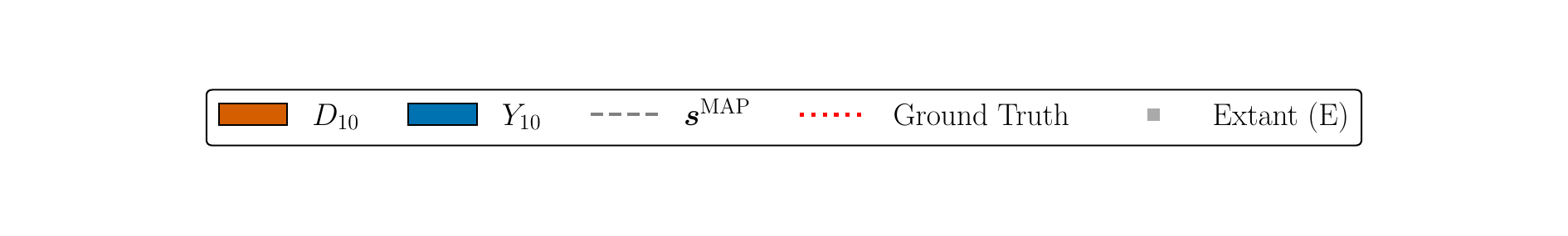}
  \caption{
      \textbf{Quantifying quality of coevolution-aware reconstruction.} 
      Box-plots of three different reconstruction quality metrics for the extant sequences, $\mathcal{N}=10$ top-ranking sequences from $\ancestralset$ ($\bm{\text{Y}}_{10}$) and $\DCAset$ ($\bm{\text{D}}_{10}$), in terms of site-independent posterior probability $\Post$. 
      Results are shown for three different roots as a function of evolutionary time $\muDCA$.
      First row shows the hamming distance to the $\GT$, with black dotted line representing the hamming distance between $\MAP$ and $\GT$.
      Second row is the ESMFold sequence average pLDDT of candidate ancestors, with the red dotted line being the pLDDT of $\GT$. Third row displays the RMSD between ESMFold structural estimates and AlphaFold accurate estimation of the GT structure. 
      Here the red dotted line displays the lower bound of RMSD obtained comparing the structural estimates of ESMFold and AlphaFold over $\GT$.
}
  \label{fig:reshuffled_asr}
\end{figure*}

\subsection{Analysis of best performing candidate ancestral sequences}\label{subsec:picking_good_DCA_ancestral_sequences}
In order to find optimal reconstructed ancestors among the coevolution-aware candidate ancestors $\DCAset$, without knowing $\GT$, we need to find a quantity to guide our choice.
Here, we choose the posterior probability $\Post$ to pick the best sequences. Interestingly, this probability seems to be an acceptable proxy for ASR quality, the probabilistic uncertainty often correlating with Hamming divergence from the GT (see SI, Fig.~S9). However, once again, this correlation is not exact and shows a marked dependency on root mutability and intra-leaf divergence -- represented by $\muDCA$ -- especially when plotted on sampled sequences from $\ancestralset$ (see SI, Fig.~S10). 

We select the $\mathcal{N}$ best sequences according to $\Post$, from either $\ancestralset$ or $\DCAset$. 
Fig.~\ref{fig:reshuffled_asr} shows the results of this selection for $\mathcal{N}=10$ (also plotted on the $E_\text{DCA}$ vs $d_\text{H}(\cdot, \GT)$ grid plot in SI, Fig.~S7, and on the PCA space of the first two Principal Components of the extant family, in SI Fig.~S8). 
Similar results over the DNA Binding Domain family are presented in SI. 
In the first row, we show the resulting Hamming distance of the $\mathcal{N}$ best picks to $\GT$, always comparing it to the distance between $\MAP$ and $\GT$ (black dotted line).
In all cases, average Hamming distance of the $10$ best DCA samples gravitates around $d_\text{H}(\MAP, \GT)$, even outperforming the MAP sequence in terms of proximity to the GT, in some cases. 
However, as shown and explained in the preceding section, site-independent samples often seem quite distant from the GT as soon as the evolutionary distance grows. 
Our model thus seems to sample acceptable ancestral sequences, producing diverse candidates that share close proximity to the GT. 

At extended evolutionary timescales, sampling noise and the stochasticity of the evolutionary process make perfectly accurate ancestral recovery fundamentally unattainable. The true ancestor should therefore be understood as a draw from a distribution over equally plausible sequences, rather than a single recoverable point. 
In this regime, the ability to sample a \textit{diverse} set of candidate ancestors becomes a virtue in itself: a broad but well-constrained ensemble reduces the risk of over-committing to a single reconstruction that may reflect inference artifacts rather than true ancestral signal. 
By preserving the site-independent posterior while imposing coevolutionary constraints, our reshuffling procedure produces candidates that explore this plausible ancestral distribution more faithfully -- and with lower risk of systematic bias -- than methods that return a single MAP estimate or unconstrained samples.

Hamming distance to the ground truth alone is insufficient to assess ASR quality, since biological function is ultimately determined by three-dimensional structure. We therefore evaluate the structural plausibility of reconstructed sequences using ESMFold~\cite{Lin2023evolutionary}, reporting the predicted Local Distance Difference Test (pLDDT) as a proxy for local structural confidence in the second row of Fig.~\ref{fig:reshuffled_asr}.

Sequences in $\DCAset$ consistently yield higher average pLDDT scores than those in $\ancestralset$, in some cases exceeding even the pLDDT of the GT (red dotted line) and of extant sequences (grey box). This suggests that incorporating coevolutionary constraints filters out sampling noise that would otherwise produce physically unrealistic residue pairings, yielding ancestors that are both closer to the GT in sequence space and structurally coherent.

To assess structural similarity more directly, we compute the RMSD between the AlphaFold~\cite{jumper2021highly} predicted structure of the GT and the ESMFold~\cite{Lin2023evolutionary} predictions for candidate ancestors in $\ancestralset$ and $\DCAset$ (third row of Fig.~\ref{fig:reshuffled_asr}). We use AlphaFold for the GT as a high-accuracy reference structure prediction, while ESMFold is used consistently across ancestral candidates to ensure a uniform and scalable evaluation of individual reconstructed sequences (i.e. without relying on multiple-sequence alignments). A lower bound is provided by the RMSD between AlphaFold and ESMFold predictions of the GT itself.

For GT1 and GT3, coevolution-aware candidates systematically achieve smaller structural deviation from the GT --- strikingly, with accuracy exceeding that of extant sequences. For GT5, however, the lower-bound RMSD is itself very large, indicating that ESMFold cannot reliably reproduce the AlphaFold structure for this root; predictions over candidate ancestors are therefore untrustworthy, likely because GT5 requires homology-based modeling that ESMFold bypasses.

Taken together, these results demonstrate that integrating coevolutionary information via DCA provides a robust framework for ASR. By using the site-independent posterior probability as a proxy for sequence quality, $\DCAset$ candidates not only remain close to the ground truth in sequence space but exhibit systematically superior structural properties --- higher pLDDT scores and reduced RMSD --- reflecting the importance of coevolutionary constraints for recovering ancestors that are both statistically plausible and biologically viable.

\section{Discussion}

This work addresses the long-standing limitation of site-independence in ancestral sequence reconstruction. By combining standard Bayesian ASR with a DCA-informed reshuffling step, we explicitly construct candidate ancestral sequences that: (i) achieve reconstruction accuracy comparable to the MAP site-independent estimate in terms of Hamming distance to the ground truth; (ii) capture the intrinsic uncertainty of the problem through a diverse ensemble rather than a single point estimate; (iii) exhibit coevolutionary scores comparable to those of natural extant sequences; and (iv) yield structural predictions closer to the known GT structure. This hybrid approach enforces coevolutionary constraints that site-independent methods ignore by construction, while getting around the intractability of summing on a phylogeny under a DCA prior.

Rigorously testing this improvement requires a benchmarking setting that is itself governed by epistatic constraints -- a condition that standard substitution models cannot satisfy by construction, as simulating under a site-independent model would suppress the very signal whose recovery is being tested. Our DCA-based forward simulator provides a biologically grounded alternative, generating evolutionary trajectories that respect residue-residue interactions and remain in the high-fitness regions of sequence space characteristic of natural protein families. The gain brought by coevolution-aware reconstruction is therefore only detectable and meaningful when the underlying data reflects true epistatic constraints, as in natural evolution or our generative model. This is not merely a technical caveat, but a conceptual result in its own right.

We started our analysis by testing site-independent ASR on \textit{in-silico} data from our coevolution-aware evolutionary simulations. 
Results show that while Maximum A Posteriori estimates often achieve low Hamming distances to the ground truth, they tend to overfit the message from the leaves, producing atypical sequences with extraordinary properties. 
Studying MAP reconstruction quality of root sequences with different mutability, encoded by a DCA-based mutational tolerance metric, we identify an evolutionary time interval in which incorporating phylogenetic correlations outperforms trivial consensus reconstruction. 
These relevant timescales are those where reconstruction is non-trivial -- avoiding the simplicity of very short divergence times -- yet remains feasible before the signal is lost to saturated mutations.
In this regime, sophisticated models demonstrate a clear advantage over simpler consensus-based approaches.

By centering the benchmarking on this relevant temporal regime, we demonstrate that integrating coevolutionary signals into the reconstruction process effectively steers ancestral candidates toward a sequence space that is both evolutionarily consistent and structurally viable. 
Our coevolution-aware ASR procedure generates ancestral sets that not only match the MAP estimate in terms of Hamming proximity to the ground truth and DCA energies -- a robust proxy for functional fitness -- but also significantly outperform site-independent posterior sampling across both metrics. 
Crucially, this superiority extends to the structural level; candidates from the coevolution-aware set exhibit both higher pLDDT scores in ESMFold structural models, indicating enhanced local folding confidence, and lower RMSD values relative to the ground truth structure. This simultaneous optimization of sequence accuracy and structural coherence suggests that coevolutionary constraints are indispensable for filtering out the sampling noise that typically plagues deep-time reconstructions, ultimately recovering ancestors that are more biologically authentic than those produced by traditional site-independent methods.

These findings have several important implications. 
First, we have consistently shown that except in trivial settings, the MAP sequence does not exactly recover the ground-truth ancestor. 
Therefore, site-independent MAP should be interpreted cautiously when drawing conclusions about function, stability, or other biochemical properties, especially since we show that the MAP sequence indirectly overfits DCA energy, leading to extraodinary thermostability and catalytic function. 
Considering ancestral reconstruction uncertainty in most regimes, we advocate to take into account the inherent stochasticity of evolutionary processes rather than to focus on a single likely reconstruction. 
This argues in favor of sampling from the posterior ancestral distribution to generate an ensemble of plausible ancestral sequences, instead of relying on a single-point estimate. 
Our results further indicate that this objective is more effectively achieved using a DCA-informed posterior, particularly in evolutionary regimes where samples drawn from the purely site-independent posterior are rarely functional or biologically meaningful.
The fact that our method achieves better predicted biological and structural properties without sacrificing proximity to the ground truth underscores the importance of coevolution in shaping protein evolution and the necessity to integrate coevolutionary signals in phylogenetic pipelines.

In this work, our method was applied to a single node in the tree. 
While this procedure is applicable to any node, a limitation of our method is that this must be done independently for each node.  
In other words, we can generate a diverse set of potential ancestors for any set of leaves but are unable to generate consistent evolutionary trajectories across the phylogeny, as is possible with site-independent Bayesian frameworks.
Overcoming this limitation would involve the development of scalable ancestral sampling schemes that maintain consistency across nodes while ensuring that each step along the trajectory remains within the functional sequence space defined by coevolutionary interactions.
Another limitation is the strong reliance on ground-truth information to quantify the accuracy and biological relevance of reconstructed sequences. 
In empirical settings, this information is unavailable. 
An improvement would be to design evaluation protocols based solely on properties of the reconstructed sequences. 
Examples are the pLDDT metric that we used above, or the difference in entropy between the leaves and the site-independent posterior distribution, which measures how much uncertainty is removed by ASR. 
Notably, adding coevolution information to the ancestral posterior reduces the entropy of the ancestral distribution (see SI, Fig.~S10), therefore constraining the sampled sequences to a more functional sequence space and reducing ancestral uncertainty. 

An important potential use of our DCA-informed ASR framework is in experimental protein resurrection. 
By providing an efficient approach to incorporate coevolutionary constraints, the method allows generating a diverse set of ancestral sequences that are more likely to be bio-physically viable and less biased than a single MAP reconstruction. 
Assessing the functional landscape, stability, and enzymatic properties of these reconstructions \textit{in vitro} will be essential to confirm how our results concretely translate for real biological systems. 
Such experiments would provide an important step to assess the relevance of coevolutionary models in phylogenetics.

\section{Methods}\label{sec:methods}
\subsection{Alignment cleaning and tree inference}
\label{method:tree_and_msa_cleaning}
For a given seed of sequences of interest, we generate a Multiple Sequence Alignment (MSA) and infer its phylogenetic tree using FastTree~\cite{priceMorgannpriceFasttree2025}. The choice of this specific inference tool instead of more precise ones like IQTree~\cite{IqtreeIqtree32025} and PhyloTree~\cite{VegPhylotreejs2025}, to cite a few, is mainly motivated by the size of the tree that needs to be inferred ($\approx 10000$ leaves), and is further described and argued in SI. Both the MSA and the tree are curated as described in SI. We make sure that the curation process improves data quality with minimal information loss (see SI, Fig.~S1).
\subsection{Inference of the DCA model}
\label{method:DCA_inference}
Naturally, incorporating coevolution signal within sequences in the ASR and benchmarking procedures requires the inference of a coevolution-aware model. 
We thus infer a coevolution generative model using the DCA method~\cite{figliuzziHowPairwisecoevolutionary2018, cocco2018inverse}.  
DCA defines a probability distribution over the sequence space of a given protein family through an energy-based model.
In particular, our DCA model assigns a probability $\DCAprob(\vect{s})$ to each amino acid sequence $\vect{s}= (s_1, \dots, s_L)$ of length $L$, such that:

\begin{equation}
    \DCAprob(\vect{s}) = \frac{1}{Z} \exp\left[ - E(\vect{s}) \right] \ ,
\end{equation}
where $E$ is the DCA energy given by
\begin{equation}
E(\vect{s})
= - \sum_{i=1}^L h_i(s_i)
  - \sum_{i < j} J_{ij}(s_i, s_j) \ ,
  \label{eq:energy}
\end{equation}
and $Z$ is the normalization constant.
The fields $h_i$ encode position-specific amino acid preferences, while the couplings $J_{ij}$ capture coevolution interactions between pairs of residues. 
Using adabmDCA~\cite{rossetAdabmDCA20Flexible2025}, we learn a model for the protein family, taking $\Data_\text{extant}$ as input. 
The inference procedure ensures that any sample from the DCA model probability distribution will reproduce the empirical single-site $f_i(a)$ and pairwise amino acid frequencies $f_{ij}(a,b)$ observed in extant sequences~\cite{cocco2018inverse}.

This unsupervised model has proved to be efficient in different tasks such as contact prediction~\cite{marks2011protein}, generation of in silico functional proteins and ribozymes~\cite{russ2020evolution, Calvanese2025generating, Lambert2024expanding, netti2026expanding}, prediction of epistatic mutability constraints~\cite{rodriguez2022epistatic, vigue2022deciphering} and the study of epistatic networks and their influence on evolutionary trajectories~\cite{de2020epistatic, bisardi2022modeling, alvarez2024vivo, DiBari2024, chen2023understanding}.

The predictive power of the model lies in its ability to disentangle direct correlations from indirect ones. Specifically, the strongest coupling parameters $J_{ij}$ serve as a robust proxy for physical structural contacts, as they capture the direct coevolutionary constraints between residues in the three-dimensional fold. Furthermore, it has been demonstrated~\cite{cocco2018inverse, russ2020evolution, Lambert2024expanding, netti2026expanding} that within a single protein family, sequences with low DCA energies exhibit a higher probability of experimental functionality. 
Conversely, sequences characterized by high DCA energies are almost certainly non-functional, suggesting that the DCA energy serves as a reliable proxy for biological viability.

The inference of parameters completely ignores the phylogenetic information, yet studies suggest that the global nature of the DCA model provides an inherent resilience to evolutionary noise. While phylogenetic correlations are known to produce spurious couplings that limit the detection of weaker structural contacts~\cite{horta2021effect}, the strongest DCA  remain remarkably robust and typically represent true physical constraints rather than historical artifacts. Furthermore, global inference methods have been shown to filter these hierarchical correlations more effectively than local statistical measures~\cite{dietler2023impact}, confirming that the model naturally distinguishes direct coevolution from deep-rooted phylogenetic signals.

\subsection{Forward evolution algorithm}
\label{method:Forward_evolution_algorithm}

To generate synthetic datasets that incorporate coevolutionary constraints, we simulate sequence evolution as a stochastic process along a rooted phylogenetic tree. 
Previous DCA-based forward simulators over independent lineages showed that this procedure models how a protein sequence diverges from an ancestral root while maintaining the structural and functional constraints captured by the DCA model, with sequences along evolutionary trajectories successfully resulting functional in experimental assays~\cite{de2020epistatic, alvarez2022novel, alvarez2024vivo}.

Starting from the root of the tree with a reference amino acid sequence $\GT$, evolution is propagated down the tree recursively from parent to child. 
For every branch connecting a parent node $u$ to a child node $v$, we compute the number of mutation attempts based on the inferred branch length $t_{uv}$. 
The total number of mutation attempts on a branch is given by $N_{mut} = \muDCA L t_{uv}$, where $L$ is the sequence length and $\muDCA$ is a global scaling factor representing the mutation rate. 
After attempting mutations for a given branch, the resulting sequence is assigned to node $v$ and serves as the template for evolution toward its own descendants.

For each mutation attempt on a branch, we perform a site-specific update using a Gibbs sampling scheme. 
To incorporate coevolution, the probability of a mutation at a specific site depends on the residues currently occupying all other positions in the sequence. 
A site $i \in \{1, \dots, L\}$ is selected uniformly at random, and a new amino acid $a$ is drawn according to the conditional probability distribution:
\begin{equation}
\label{eq:cond_prob_SI}
\DCAprob(s_i = a \mid \vect{s}_{\setminus i}) = \frac{1}{Z_i} \exp\left( h_i(a) + \sum_{j \neq i} J_{ij}(a, s_j) \right),
\end{equation}
where $Z_i$ is a normalization constant.

This update scheme ensures that the forward simulation is coevolution-aware. 
The term $h_i(a)$ represents the site-specific preference for amino acid $a$, while the term $\sum_{j \neq i} J_{ij}(a, s_j)$ accounts for the interaction between the proposed amino acid $a$ at site $i$ and the current residues $s_j$ at all other sites $j$. 
Mutations that are incompatible with the current background sequence are statistically suppressed, effectively mimicking the effect of natural selection preserving the protein's structural integrity. 

An algorithmic pseudo-code describing the forward evolutionary process is given as follows:

\begin{enumerate}
    \item \textbf{Initialize:} Set root sequence $\vect{s}_R = \GT$.
    \item \textbf{Recursive Step:} For each node $u$ with child $v$:
    \begin{itemize}
        \item Set $\vect{s}_v = \vect{s}_u$.
        \item Compute $N_{mut} = \text{round}(\muDCA L t_{uv})$ where ``round'' is the standard stochastic rounding.
        \item \textbf{Loop} $k=1$ to $N_{mut}$:
        \begin{enumerate}
            \item Select site $i \in \{1, \dots, L\}$ at random.
            \item Update $s_{v,i}$ by sampling $a \sim \DCAprob(a \mid \vect{s}_{v \setminus i})$.
        \end{enumerate}
        \item Repeat for all descendants.
    \end{itemize}
    \item \textbf{Output:} Collect sequences at leaf nodes as $\Data$.
\end{enumerate}

It is important to note that all evolutionary distances reported in figures are functions of $ \muDCA$. 
This parameter is distinct from the mutation rate $\mu$ inferred under the site-independent transition model used for ASR (see Section Methods~\ref{method:site_independent_ASR}). The relationship between $\mu$ and $\muDCA$ is shown in SI (Fig.~S4). 

\subsection{Choice of root sequences}\label{method:choice_of_root_sequences}
In line with previous studies~\cite{vigue2023predicting}, we select the $\GT$ based on their Context-Dependent Entropy (CDE)~\cite{vigueDecipheringPolymorphism611572022}, a mutability metric derived from DCA (see SI). 
CDE quantifies the effective diversity of amino acids that are compatible with a given site, conditioned on the rest of the sequence under the DCA model. 
In other words, it captures how strongly a position is constrained by coevolutionary interactions with other residues in the sequence.
Low CDE values indicate that only a limited set of residues is compatible with the sequence context, reflecting strong coevolutionary constraints and low mutability. In contrast, high CDE values correspond to more permissive contexts, where a broader range of residues is tolerated, indicating higher mutability. 

For a given sequence $\vect{s}$ and site $i$, the context-dependent probability of observing amino acid $a$ at that site, conditioned on the rest of the sequence $\vect{s}_{\setminus i}$, is defined in Eq.~\eqref{eq:cond_prob_SI}.
The context-dependent entropy is then defined as the Shannon entropy of this conditional distribution,
\begin{equation}
\begin{aligned}
\text{CDE}^{(i)}(\vect{s})
&= - \sum_{a \in \Alphabet}
\DCAprob(s_i = a \mid \vect{s}_{\setminus i}) \\
&\quad \times
\log_2 \DCAprob(s_i = a \mid \vect{s}_{\setminus i}).
\end{aligned}
\end{equation}

To be quantitative, $2^{\text{CDE}^i(\vect{s})}$ is roughly the amount of amino acids tolerated in position $i$ given the residues present in other positions of $\vect{s}$.
This means that if $\text{CDE}^i = 0$ only one amino acid or gap can occupy position $i$, whereas if $\text{CDE}^i = 4.39$ all 21 amino acids or gap are tolerated in position $i$.
By averaging this quantity over all sites, we can define a sequence-level context-dependent entropy as 
\begin{equation}
\text{CDE}(\vect{s}) = \frac{1}{L} \sum_{i=1}^{L} \text{CDE}^{(i)}(\vect{s}).
\label{eq::CDE}
\end{equation}
It has been shown~\cite{Rossi_2025} that the speed of evolutionary divergence from a starting sequence is highly contingent on said sequence mutability quantified by CDE, revealing the crucial role of sequence context in evolutionary timescales.

Consequently, we choose to simulate forward evolution starting from 5 different extant $\beta$-lactamase sequences $\GTindexed{1}, \cdots, \GTindexed{5}$ at the root. These starting sequences were selected in a wide range of $\text{CDE}$ under the DCA model (see SI, Fig.~S3). 
This variability in mutational tolerance of the root sequences then allows us to benchmark the performance of reconstruction algorithms in diverse evolutionary contexts. 
From each root, sequences were sampled down the tree as described above; the resulting leaves $\Data$ were stored for analysis, for values of $\muDCA$ ranging from 0.1 to 3000. 

\subsection{Site-independent ancestral sequence reconstruction}\label{method:site_independent_ASR} 

The sequence data $\mathcal{D}$ obtained from the forward coevolution-aware simulator are used to reconstruct the ancestral root sequence $\GT$. 
We perform ancestral sequence reconstruction (ASR) using the site-independent likelihood framework established by Yang~\cite{yangNewMethodInference1995}, which builds upon the Felsenstein pruning algorithm~\cite{felsensteinEvolutionaryTreesDNA1981}. More specifically, we seek the likelihood of a given sequence $\mathbf{s}$ at the root of the tree, denoted as $\LikRoot(\mathbf{s} \mid \Data, \Tree, \vect{\pi}, \mu)$, given the leaf sequences $\Data$, the rooted phylogenetic tree $\Tree$, and the parameters $(\vect{\pi}, \mu)$ of a reversible site-independent transition model that we will specify later in this section.

Under the simplifying assumption of site independence, the likelihood factorizes over sites such that for any root sequence $\mathbf{s}$, we have
\begin{equation}\label{eq:likelihood_factorization}
\mathcal{L}_R(\mathbf{s} \mid \Data, \Tree, \vect{\pi}, \mu) = \prod_{i=1}^L \mathcal{L}_R^{(i)}(s_i \mid \DataVec{i}, \Tree, \pi^{(i)}, \mu)
\end{equation}
where $\DataVec{i} = (s_i^1, \dots, s_i^N)$ represents the states observed at site $i$ across the $N$ leaf sequences, and $\pi^{(i)}$ is the site-wise stationary distribution.

We define the site-independent transition process that is generally needed to perform the reconstruction. 
This simple model assigns a transition probability at site $i$ from state $s_u$ at node $u$ to state $s_v$ at child node $v$ via a continuous-time Markov process:
\begin{equation}
P^{(i)}(s_u \rightarrow s_v \mid t_{uv}) = e^{-\mu t_{uv}}\delta_{s_u s_v} + (1 - e^{-\mu t_{uv}})\pi^{(i)}(s_v)
\label{eq::site_ind}
\end{equation}
with $t_{uv}$ representing the evolutionary time and $\mu$ a global mutation rate which is homogeneous across tree branches and is different from $\muDCA$ of the coevolution-aware forward simulator. 
Here, the stationary distributions $\pi^{(i)}$ are set to the empirical amino acid frequencies observed in the leaves. 

To compute the site-wise likelihood $\mathcal{L}_R^{(i)}$ efficiently, we propagate conditional likelihoods $\mathcal{L}_u^{(i)}(a)$ from the leaves to the root.
At the leaves, the likelihood is initialized as $\mathcal{L}_{\text{leaf } n}^{(i)}(a) = \delta_{a, s_i^n}$, and for any internal node $v$, it is computed recursively as
\begin{equation}
\mathcal{L}_v^{(i)}(a) = \prod_{c \in \text{children}(v)} \left( \sum_{b \in \mathcal{A}} P^{(i)}(a \rightarrow b \mid t_{vc}) \mathcal{L}_c^{(i)}(b) \right).
\end{equation}
The final site-wise posterior distribution at the root is then given by
\begin{equation}\label{eq:posterior_ancestral_site_i}
\postpersite(a \mid \DataVec{i}, \mathcal{T}, \pi^{(i)}, \mu) = \frac{\pi^{(i)}(a) \mathcal{L}_R^{(i)}(a)}{\sum_{b \in \mathcal{A}} \pi^{(i)}(b) \mathcal{L}_R^{(i)}(b)}.
\end{equation}

Sampling from $\postpersite$ produces an alignment of candidate site-independent ancestors $\ancestralset$. We also compute the Maximum A Posteriori (MAP) estimate $\MAP$, which assigns at each site the state with the largest posterior probability as 

\begin{equation}
    \MAPsite{i} = \postpersite(a \mid \DataVec{i}, \mathcal{T}, \pi^{(i)}, \mu)
\end{equation}
with $\Alphabet$ the alphabet of possible characters (amino acids or nucleotides plus the gap symbol).  

Rather than inferring the global mutation rate $\mu$ via maximization of the phylogenetic likelihood, we adopt a calibration procedure based on pairwise sequence divergence. We compute the normalized Hamming distances between leaf pairs and plot them as a function of the corresponding phylogenetic distances $t$ separating them on the tree $\mathcal{T}$. 
Considering that our site-independent propagator is time-reversible, $t$ is taken as the sum of the branch lengths over the shortest path connecting these two leaves.
Under our site-independent transition model, the expected Hamming distance as a function of tree distance $t$ follows the exponential saturation form $\mathbb{E}[d_H(t)] = a(1 - e^{-\mu t})$, where $a$ accounts for the finite alphabet size and residue frequencies. 
The parameters $(a, \mu)$ are estimated via nonlinear least-squares fitting, providing a robust and computationally efficient calibration for the global mutation rate $\mu$ (see SI, Fig.~S4).

\subsection{Coevolution-aware ancestral sequence reconstruction}\label{method: co_evolution_aware_ASR}

The previous reconstruction framework assumes site independence, hence fundamentally neglecting coevolutionary interactions between residues. 
This limitation motivates to extend ASR beyond site-independent models, by explicitly incorporating a coevolution signal. 
A principled approach would be to perform exact inference on the phylogenetic tree using max-product algorithms under a DCA prior. 
However, this would require summing over all possible ancestral sequences, whose state space grows exponentially with sequence length.

To circumvent this limitation, we propose an alternative strategy that preserves tractability while introducing coevolutionary constraints. 
Our approach consists of the following steps:

\begin{enumerate}
    \item Independent Site Reconstruction: We first perform the site-independent ancestral reconstruction to obtain the site-wise posterior distributions $\postpersite$ at the root of the tree.
    
    \item Initial Sampling: A set of $M=1000$ candidate ancestral sequences is sampled from the product posterior distribution
    $\Post=\prod_{i=1}^L \postpersite$. This yields a site-independent alignment of candidate ancestors $\ancestralset=(s^\alpha_{\text{R},i})^{\alpha \in \{1, \cdots, M\}}_{i \in \{1, \cdots, L\}} $, where columns represent sites and rows represent individual sequences. The choice of $M$ is a heuristic balance, providing sufficient diversity while remaining computationally tractable for the subsequent optimization.
    
    \item Coevolutionary Optimization: The alignment is refined through a column-wise amino acid swap procedure to minimize the DCA score. In each iteration:
    \begin{enumerate}
        \item A site $i$ is selected with probability proportional to its Shannon entropy $\Gamma_i$ (calculated from the frequencies in $\ancestralset$), ensuring the optimization focuses on non-conserved positions.
        \item Two sequences, $s_R^\alpha$ and $s_R^\beta$, are randomly chosen for an amino acid swap at site $i$, resulting in trial sequences $\hat{s}_R^\alpha$ and $\hat{s}_R^\beta$.
        \item The swap is accepted with a Metropolis probability $\min(1, e^{-\Delta E_{i,\alpha\beta}/T})$, where $\Delta E_{i,\alpha\beta} = [ E(\hat{s}_R^\alpha) + E(\hat{s}_R^\beta) ] - [ E(s_R^\alpha) + E(s_R^\beta) ]$. We utilize $T=0.2$ as an effective temperature to regulate the selection stringency.
    \end{enumerate}
    
    \item Termination: The procedure stops heuristically after $2^{M \cdot \Gamma / 2}$ attempted moves, where $\Gamma = \sum_{i=1}^L \Gamma_i$ is the total entropy of the alignment. For site $i$, we define $\Gamma_i=-\sum_{a \in \Alphabet} f_i(a)\log(f_i(a))$ as the site-wise entropy of the amino acid distribution of the ancestral set $\ancestralset$, taking the single-point frequencies $f_i$ of the amino acids as a probability distribution per site.
    The procedure gives a coevolution-aware alignment of candidate ancestors $\DCAset$.
\end{enumerate}

In summary, this  procedure preserves the per-site amino acid probabilities and hence the probability of the ancestral sequences under the independent-site model.
As a consequence, $\DCAset $ satisfies the single site ancestral $\postpersite$  that has been used to sample the site-independent candidate ancestors $\ancestralset$,  while minimizing the DCA score, i.e. maximizing the global likelihood of the alignment $\DCAset$ under the DCA model. 
This procedure is stochastic, hence different realizations starting from the same $\ancestralset$ will produce different $\DCAset$.

\acknowledgements{We thank Sabrina Cotogno, Edwin Rodriguez-Horta, Jeanne Trinquier, and Barthélémy Meynard-Piganeau for their help in early stages of the project, Lorenzo Rosset for his help on parallelization and speed management, Francesco Calvanese for providing useful pieces of code, Guillaume Achaz, Nicolas Lartillot and Joseph Thornton for providing useful feedback. Part of this work was performed using HPC resources from GENCI--IDRIS (Grant 2023-[AD011014914R1]).
This research has been supported by the AMX grant of Ecole Polytechnique de Paris, by the Fondation pour la Recherche Médicale, reference number FDT202604051476, by the first FIS (Italian Science Fund) 2021 funding scheme (FIS783 - SMaC - Statistical Mechanics and Complexity) from MUR, Italian Ministry of University and Research and from the PRIN funding scheme (2022LMHTET - Complexity, disorder and fluctuations: spin glass physics and beyond) from MUR, Italian Ministry of University and Research, with additional support from the European REA, Marie Sklodowska Curie Actions, grant agreement no. 101131463 (SIMBAD).}

\bibliography{ASR}

@article{netti2026expanding,
  title={Expanding functional protein sequence space using high entropy generative models},
  author={Netti, Roberto and Hinds, Emily and Calvanese, Francesco and Ranganathan, Rama and Weigt, Martin and Zamponi, Francesco},
  journal={arXiv preprint arXiv:2605.03578},
  year={2026}
}

@article{eickRobustnessReconstructedAncestral2017,
  title = {Robustness of {{Reconstructed Ancestral Protein Functions}} to {{Statistical Uncertainty}}},
  author = {Eick, Geeta N. and Bridgham, Jamie T. and Anderson, Douglas P. and Harms, Michael J. and Thornton, Joseph W.},
  year = {2017},
  month = feb,
  journal = {Molecular Biology and Evolution},
  volume = {34},
  number = {2},
  pages = {247--261},
  issn = {1537-1719},
  doi = {10.1093/molbev/msw223},
  langid = {english},
  pmcid = {PMC6095102},
  pmid = {27795231}
}

@article{de2014empirical,
  title={Empirical fitness landscapes and the predictability of evolution},
  author={De Visser, J Arjan GM and Krug, Joachim},
  journal={Nature Reviews Genetics},
  volume={15},
  number={7},
  pages={480--490},
  year={2014},
  publisher={Nature Publishing Group UK London},
  doi={10.1038/nrg3744}
}

@article{starr2016epistasis,
  title={Epistasis in protein evolution},
  author={Starr, Tyler N and Thornton, Joseph W},
  journal={Protein science},
  volume={25},
  number={7},
  pages={1204--1218},
  year={2016},
  publisher={Wiley Online Library},
  doi={10.1002/pro.2897}
}

@article{johnson2023epistasis,
  title={Epistasis and evolution: recent advances and an outlook for prediction},
  author={Johnson, Milo S and Reddy, Gautam and Desai, Michael M},
  journal={BMC biology},
  volume={21},
  number={1},
  pages={120},
  year={2023},
  publisher={Springer},
  doi={10.1186/s12915-023-01585-3}
}

@article {DiBari2026modeling,
	author = {Di Bari, Leonardo and Mora, Thierry and Pagnani, Andrea and Walczak, Aleksandra M. and Zamponi, Francesco and Rossi, Saverio},
	title = {Modeling Protein Evolution via Generative Inference From Monte Carlo Chains to Population Genetics},
	elocation-id = {2026.02.09.704757},
	year = {2026},
	doi = {10.64898/2026.02.09.704757},
	publisher = {Cold Spring Harbor Laboratory},
	journal = {bioRxiv}
}

@article{pagnani2025generative,
  title={Generative continuous time model reveals epistatic signatures in protein evolution},
  author={Pagnani, Andrea and Barrat-Charlaix, Pierre},
  journal={bioRxiv},
  pages={2025--09},
  year={2025},
  publisher={Cold Spring Harbor Laboratory}
}

@misc{Koehl2026deep,
   author = {Antoine Koehl and Sebastian Prillo and Matthew Liu and Junhao Xiong and Lillian Weng and David F. Savage and Yun S. Song},
   doi = {10.64898/2026.02.19.706898},
   institution = {bioRxiv},
   month = {2},
   title = {Deep models of protein evolution in time generate realistic evolutionary trajectories and functional proteins},
   year = {2026}
}

@article{Hie2022evolutionary,
   author = {Brian L. Hie and Kevin K. Yang and Peter S. Kim},
   doi = {10.1016/j.cels.2022.01.003},
   issn = {24054720},
   issue = {4},
   journal = {Cell Systems},
   month = {4},
   pages = {274-285.e6},
   pmid = {35120643},
   publisher = {Cell Press},
   title = {Evolutionary velocity with protein language models predicts evolutionary dynamics of diverse proteins},
   volume = {13},
   year = {2022}
}

@misc{Gorstein2025ancestral,
   author = {Evan Gorstein and Mengze Tang and Hailey Bruzzone and Claudia Solís-Lemus},
   doi = {10.1101/2025.11.19.689264},
   institution = {bioRxiv},
   month = {11},
   title = {Ancestral Sequences Cannot be Accurately Reconstructed via Interpolation in a Variational Autoencoder’s Latent Space},
   year = {2025}
}

@article{Ding2019deciphering,
   author = {Xinqiang Ding and Zhengting Zou and Charles L. Brooks},
   doi = {10.1038/s41467-019-13633-0},
   issn = {20411723},
   issue = {1},
   journal = {Nature Communications},
   month = {12},
   pmid = {31822668},
   publisher = {Nature Research},
   title = {Deciphering protein evolution and fitness landscapes with latent space models},
   volume = {10},
   year = {2019}
}

@inproceedings{vaswani2017attention,
 author = {Vaswani, Ashish and Shazeer, Noam and Parmar, Niki and Uszkoreit, Jakob and Jones, Llion and Gomez, Aidan N and Kaiser, \L ukasz and Polosukhin, Illia},
 booktitle = {Advances in Neural Information Processing Systems},
 editor = {I. Guyon and U. Von Luxburg and S. Bengio and H. Wallach and R. Fergus and S. Vishwanathan and R. Garnett},
 pages = {},
 publisher = {Curran Associates, Inc.},
 title = {Attention is All you Need},
 volume = {30},
 year = {2017}
}

@article{weigt2009identification,
  title={Identification of direct residue contacts in protein--protein interaction by message passing},
  author={Weigt, Martin and White, Robert A and Szurmant, Hendrik and Hoch, James A and Hwa, Terence},
  journal={Proceedings of the National Academy of Sciences},
  volume={106},
  number={1},
  pages={67--72},
  year={2009},
  publisher={National Academy of Sciences},
  doi={10.1073/pnas.0805923106}
}

@article{ly2022alisim,
  title={AliSim: a fast and versatile phylogenetic sequence simulator for the genomic era},
  author={Ly-Trong, Nhan and Naser-Khdour, Suha and Lanfear, Robert and Minh, Bui Quang},
  journal={Molecular biology and evolution},
  volume={39},
  number={5},
  pages={msac092},
  year={2022},
  publisher={Oxford University Press},
  doi={10.1093/molbev/msac092}
}

@article{fletcher2009indelible,
  title={INDELible: a flexible simulator of biological sequence evolution},
  author={Fletcher, William and Yang, Ziheng},
  journal={Molecular biology and evolution},
  volume={26},
  number={8},
  pages={1879--1888},
  year={2009},
  publisher={Oxford University Press},
  doi={10.1093/molbev/msp098}
}

@article{felsensteinEvolutionaryTreesDNA1981,
  title = {Evolutionary Trees from {{DNA}} Sequences: A Maximum Likelihood Approach},
  shorttitle = {Evolutionary Trees from {{DNA}} Sequences},
  author = {Felsenstein, J.},
  year = {1981},
  journal = {Journal of Molecular Evolution},
  volume = {17},
  number = {6},
  pages = {368--376},
  issn = {0022-2844},
  doi = {10.1007/BF01734359},
  langid = {english},
  pmid = {7288891}
}

@article{pauling_1963,
  title   = {Chemical Paleogenetics. Molecular ``Restoration Studies'' of Extinct Forms of Life},
  author  = {Pauling, Linus and Zuckerkandl, Emile and Henriksen, Thormod and L{\"o}vstad, Rolf},
  journal = {Acta Chemica Scandinavica},
  year    = {1963},
  volume  = {17},
  number  = {Supplement},
  pages   = {9--16},
  doi     = {10.3891/acta.chem.scand.17s-0009}
}

@article{prakineeAncestralSequenceReconstruction2024,
  title = {Ancestral {{Sequence Reconstruction}} for {{Designing Biocatalysts}} and {{Investigating}} Their {{Functional Mechanisms}}},
  author = {Prakinee, Kridsadakorn and Phaisan, Suppalak and Kongjaroon, Sirus and Chaiyen, Pimchai},
  year = {2024},
  month = dec,
  journal = {JACS Au},
  volume = {4},
  number = {12},
  pages = {4571--4591},
  publisher = {American Chemical Society},
  doi = {10.1021/jacsau.4c00653},
  urldate = {2025-09-30}
}

@misc{priceMorgannpriceFasttree2025,
  title = {Morgannprice/Fasttree},
  author = {Price, Morgan},
  year = {2025},
  month = sep,
  urldate = {2025-09-30},
  copyright = {GPL-3.0}
}

@article{spenceAncestralSequenceReconstruction2021,
  title={Ancestral sequence reconstruction for protein engineers},
  author={Spence, Matthew A and Kaczmarski, Joe A and Saunders, Jake W and Jackson, Colin J},
  journal={Current opinion in structural biology},
  volume={69},
  pages={131--141},
  year={2021},
  publisher={Elsevier},
  doi= {{10.1016/j.sbi.2021.04.001}}
}

@article{trostSimulationsSequenceEvolution2024,
  title = {Simulations of {{Sequence Evolution}}: {{How}} ({{Un}})Realistic {{They Are}} and {{Why}}},
  shorttitle = {Simulations of {{Sequence Evolution}}},
  author = {Trost, Johanna and Haag, Julia and H{\"o}hler, Dimitri and Jacob, Laurent and Stamatakis, Alexandros and Boussau, Bastien},
  editor = {Crandall, Keith},
  year = {2024},
  month = jan,
  journal = {Molecular Biology and Evolution},
  volume = {41},
  number = {1},
  pages = {msad277},
  issn = {0737-4038, 1537-1719},
  doi = {10.1093/molbev/msad277},
  urldate = {2025-09-30},
  copyright = {https://creativecommons.org/licenses/by-nc/4.0/},
  langid = {english}
}

@article{yangNewMethodInference1995,
  title = {A {{New Method}} of {{Inference}} of {{Ancestral Nucleotide}} and {{Amino Acid Sequences}}},
  author = {Yang, Z. and Kumar, S. and Nei, M.},
  year = {1995},
  month = dec,
  journal = {Genetics},
  volume = {141},
  number = {4},
  pages = {1641--1650},
  issn = {0016-6731},
  doi = {10.1093/genetics/141.4.1641},
  urldate = {2025-10-01},
  pmcid = {PMC1206894},
  pmid = {8601501}
}

@article{figliuzziHowPairwiseCoevolutionary2018,
  title = {How {{Pairwise Coevolutionary Models Capture}} the {{Collective Residue Variability}} in {{Proteins}}?},
  author = {Figliuzzi, Matteo and {Barrat-Charlaix}, Pierre and Weigt, Martin},
  year = {2018},
  month = apr,
  journal = {Molecular Biology and Evolution},
  volume = {35},
  number = {4},
  pages = {1018--1027},
  issn = {0737-4038},
  doi = {10.1093/molbev/msy007},
  urldate = {2025-10-01}
}

@article{whelanGeneralEmpiricalModel2001a,
  title = {A General Empirical Model of Protein Evolution Derived from Multiple Protein Families Using a Maximum-Likelihood Approach},
  author = {Whelan, S. and Goldman, N.},
  year = {2001},
  month = may,
  journal = {Molecular Biology and Evolution},
  volume = {18},
  number = {5},
  pages = {691--699},
  issn = {0737-4038},
  doi = {10.1093/oxfordjournals.molbev.a003851},
  langid = {english},
  pmid = {11319253}
}

@misc{VegPhylotreejs2025,
  title = {Veg/Phylotree.Js},
  year = {2025},
  month = sep,
  urldate = {2025-10-01},
  copyright = {MIT},
  howpublished = {iGEM/UCSD evolutionary biology and bioinformatics group}
}

@misc{IqtreeIqtree32025,
  title = {Iqtree/Iqtree3},
  year = {2025},
  month = oct,
  urldate = {2025-10-01},
  copyright = {GPL-2.0},
  howpublished = {iqtree}
}

@online{rossetAdabmDCA20Flexible2025, 
  title = {{{adabmDCA}} 2.0 -- a Flexible but Easy-to-Use Package for {{Direct Coupling Analysis}}},
  author = {Rosset, Lorenzo and Netti, Roberto and Muntoni, Anna Paola and Weigt, Martin and Zamponi, Francesco},
  date = {2025-01-30},
  eprint = {2501.18456},
  eprinttype = {arXiv},
  eprintclass = {q-bio},
  doi = {10.48550/arXiv.2501.18456},
  url = {http://arxiv.org/abs/2501.18456},
  urldate = {2025-10-09},
  pubstate = {prepublished}
}

@article{weissLastUniversalCommon2018,
  title = {The Last Universal Common Ancestor between Ancient {{Earth}} Chemistry and the Onset of Genetics},
  author = {Weiss, Madeline C. and Preiner, Martina and Xavier, Joana C. and Zimorski, Verena and Martin, William F.},
  year = 2018,
  month = aug,
  journal = {PLoS Genetics},
  volume = {14},
  number = {8},
  pages = {e1007518},
  issn = {1553-7390},
  doi = {10.1371/journal.pgen.1007518},
  urldate = {2026-02-04},
  pmcid = {PMC6095482},
  pmid = {30114187}
}

@article{dalenDeeptimePaleogenomicsLimits2023,
  title = {Deep-Time Paleogenomics and the Limits of {{DNA}} Survival},
  author = {Dal{\'e}n, Love and Heintzman, Peter D. and Kapp, Joshua D. and Shapiro, Beth},
  year = 2023,
  month = oct,
  journal = {Science (New York, N.Y.)},
  volume = {382},
  number = {6666},
  pages = {48--53},
  issn = {0036-8075},
  doi = {10.1126/science.adh7943},
  urldate = {2026-02-08},
  pmcid = {PMC10586222},
  pmid = {37797036}
}

@incollection{jukesEvolutionProteinMolecules1969,
  title = {Evolution of {{Protein Molecules}}},
  booktitle = {Mammalian {{Protein Metabolism}}},
  author = {Jukes, Thomas H. and Cantor, Charles R.},
  year = 1969,
  pages = {21--132},
  publisher = {Elsevier},
  doi = {10.1016/B978-1-4832-3211-9.50009-7},
  urldate = {2026-02-08},
  isbn = {978-1-4832-3211-9},
  langid = {english}
}

@incollection{Dayhoff1978Model,
  author    = {Dayhoff, Margaret O. and Schwartz, Robert M. and Orcutt, Brian C.},
  title     = {A model of evolutionary change in proteins},
  booktitle = {Atlas of Protein Sequence and Structure},
  editor    = {Dayhoff, Margaret O.},
  volume    = {5},
  number    = {Supplement 3},
  pages     = {345--352},
  publisher = {National Biomedical Research Foundation},
  address   = {Washington, DC},
  year      = {1978}
}

@article{horta2021effect,
    doi = {10.1371/journal.pcbi.1008957},
    author = {Rodriguez Horta, Edwin AND Weigt, Martin},
    journal = {PLOS Computational Biology},
    publisher = {Public Library of Science},
    title = {On the effect of phylogenetic correlations in coevolution-based contact prediction in proteins},
    year = {2021},
    month = {05},
    volume = {17},
    pages = {1-17},
    number = {5},
}

@article{jumper2021highly,
title = "Highly accurate protein structure prediction with AlphaFold",
author = "John Jumper and Richard Evans and Alexander Pritzel and Tim Green and Michael Figurnov and Olaf Ronneberger and Kathryn Tunyasuvunakool and Russ Bates and Augustin {\v Z}{\'i}dek and Anna Potapenko and Alex Bridgland and Clemens Meyer and Kohl, \{Simon A.A.\} and Ballard, \{Andrew J.\} and Andrew Cowie and Bernardino Romera-Paredes and Stanislav Nikolov and Rishub Jain and Jonas Adler and Trevor Back and Stig Petersen and David Reiman and Ellen Clancy and Michal Zielinski and Martin Steinegger and Michalina Pacholska and Tamas Berghammer and Sebastian Bodenstein and David Silver and Oriol Vinyals and Senior, \{Andrew W.\} and Koray Kavukcuoglu and Pushmeet Kohli and Demis Hassabis",
note = "Publisher Copyright: {\textcopyright} 2021, The Author(s).",
year = "2021",
month = aug,
day = "26",
doi = "10.1038/s41586-021-03819-2",
language = "English",
volume = "596",
pages = "583--589",
journal = "Nature",
issn = "0028-0836",
publisher = "Nature Research",
number = "7873",
}

@article{
Lin2023evolutionary,
author = {Zeming Lin  and Halil Akin  and Roshan Rao  and Brian Hie  and Zhongkai Zhu  and Wenting Lu  and Nikita Smetanin  and Robert Verkuil  and Ori Kabeli  and Yaniv Shmueli  and Allan dos Santos Costa  and Maryam Fazel-Zarandi  and Tom Sercu  and Salvatore Candido  and Alexander Rives },
title = {Evolutionary-scale prediction of atomic-level protein structure with a language model},
journal = {Science},
volume = {379},
number = {6637},
pages = {1123-1130},
year = {2023},
doi = {10.1126/science.ade2574},
}

@article{dietler2023impact,
  title={Impact of phylogeny on structural contact inference from protein sequence data},
  author={Dietler, Nicola and Lupo, Umberto and Bitbol, Anne-Florence},
  journal={Journal of The Royal Society Interface},
  volume={20},
  number={199},
  pages={20220707},
  year={2023},
  publisher={The Royal Society},
  doi={10.1098/rsif.2022.0707}
}

@article{jonesRapidGenerationMutation1992,
  title = {The Rapid Generation of Mutation Data Matrices from Protein Sequences},
  author = {Jones, David T. and Taylor, William R. and Thornton, Janet M.},
  year = 1992,
  month = jun,
  journal = {Bioinformatics},
  volume = {8},
  number = {3},
  pages = {275--282},
  issn = {1367-4803},
  doi = {10.1093/bioinformatics/8.3.275},
  urldate = {2026-02-08}
}

@article{pierre_arDCA,
  title = {Reconstruction of Ancestral Protein Sequences Using Autoregressive Generative Models},
  author = {De Leonardis, Matteo and Pagnani, Andrea and Barrat-Charlaix, Pierre},
  journal = {Molecular Biology and Evolution},
  year = {2025},
  volume = {42},
  number = {4},
  pages = {msaf070},
  month = {Apr},
  doi = {10.1093/molbev/msaf070},
  issn = {1537-1719}
}

@article{evansBroadcastingTreesIsing2000,
  title = {Broadcasting on {{Trees}} and the {{Ising Model}}},
  author = {Evans, William and Kenyon, Claire and Peres, Yuval and Schulman, Leonard J.},
  year = 2000,
  journal = {The Annals of Applied Probability},
  volume = {10},
  number = {2},
  eprint = {2667156},
  eprinttype = {jstor},
  pages = {410--433},
  langid = {english}
}

@article{thornton2004resurrecting,
  title={Resurrecting ancient genes: experimental analysis of extinct molecules},
  author={Thornton, Joseph W},
  journal={Nature Reviews Genetics},
  volume={5},
  number={5},
  pages={366--375},
  year={2004},
  publisher={Nature Publishing Group UK London},
  doi={10.1038/nrg1324}
}

@article{harms2010analyzing,
  title={Analyzing protein structure and function using ancestral gene reconstruction},
  author={Harms, Michael J and Thornton, Joseph W},
  journal={Current opinion in structural biology},
  volume={20},
  number={3},
  pages={360--366},
  year={2010},
  publisher={Elsevier},
  doi={10.1016/j.sbi.2010.03.005}
}

@book{liberles2007ancestral,
  title={Ancestral sequence reconstruction},
  author={Liberles, David A},
  year={2007},
  publisher={OUP Oxford},
  doi={10.1093/acprof:oso/9780199299188.001.0001}
}

@article{zakas2017enhancing,
  title={Enhancing the pharmaceutical properties of protein drugs by ancestral sequence reconstruction},
  author={Zakas, Philip M and Brown, Harrison C and Knight, Kristopher and Meeks, Shannon L and Spencer, H Trent and Gaucher, Eric A and Doering, Christopher B},
  journal={Nature biotechnology},
  volume={35},
  number={1},
  pages={35--37},
  year={2017},
  publisher={Nature Publishing Group US New York},
  doi={10.1038/nbt.3677}
}

@article{
DiBari2024,
author = {Leonardo Di Bari  and Matteo Bisardi  and Sabrina Cotogno  and Martin Weigt  and Francesco Zamponi },
title = {Emergent time scales of epistasis in protein evolution},
journal = {Proceedings of the National Academy of Sciences},
volume = {121},
number = {40},
pages = {e2406807121},
year = {2024},
doi = {10.1073/pnas.2406807121},
URL = {https://www.pnas.org/doi/abs/10.1073/pnas.2406807121},
eprint = {https://www.pnas.org/doi/pdf/10.1073/pnas.2406807121}}

@article{vigue2023predicting,
  title={Predicting the effect of mutations to investigate recent events of selection across 60,472 Escherichia coli strains},
  author={Vigu{\'e}, Lucile and Tenaillon, Olivier},
  journal={Proceedings of the National Academy of Sciences},
  volume={120},
  number={31},
  pages={e2304177120},
  year={2023},
  publisher={National Acad Sciences},
  doi={10.1073/pnas.2304177120}
}

@article{domingo2019causes,
  title={The causes and consequences of genetic interactions (epistasis)},
  author={Domingo, J{\'u}lia and Baeza-Centurion, Pablo and Lehner, Ben},
  journal={Annual review of genomics and human genetics},
  volume={20},
  pages={433--460},
  year={2019},
  publisher={Annual Reviews},
  doi={10.1146/annurev-genom-083118-014857}
}

@article{chen2023understanding,
  title={Understanding epistatic networks in the B1 $\beta$-lactamases through coevolutionary statistical modeling and deep mutational scanning},
  author={Chen, John Z and Bisardi, Matteo and Lee, Dongkyu and Cotogno, Sabrina and Zamponi, Francesco and Weigt, Martin and Tokuriki, Nobuhiko},
  journal={Nature communications},
  volume={15},
  number={1},
  pages={8441},
  year={2024},
  publisher={Nature Publishing Group UK London},
  doi={10.1038/s41467-024-54714-z}
}

@article{cocco2018inverse,
  title={Inverse statistical physics of protein sequences: a key issues review},
  author={Cocco, Simona and Feinauer, Christoph and Figliuzzi, Matteo and Monasson, R{\'e}mi and Weigt, Martin},
  journal={Reports on Progress in Physics},
  volume={81},
  number={3},
  pages={032601},
  year={2018},
  publisher={IOP Publishing},
  doi={10.1088/1361-6633/aa9965}
}

@article{russ2020evolution,
  title={An evolution-based model for designing chorismate mutase enzymes},
  author={Russ, William P and Figliuzzi, Matteo and Stocker, Christian and Barrat-Charlaix, Pierre and Socolich, Michael and Kast, Peter and Hilvert, Donald and Monasson, Remi and Cocco, Simona and Weigt, Martin and others},
  journal={Science},
  volume={369},
  number={6502},
  pages={440--445},
  year={2020},
  publisher={American Association for the Advancement of Science},
  doi={10.1126/science.aba3304}
}

@article{de2020epistatic,
  title={Epistatic contributions promote the unification of incompatible models of neutral molecular evolution},
  author={de la Paz, Jose Alberto and Nartey, Charisse M and Yuvaraj, Monisha and Morcos, Faruck},
  journal={Proceedings of the National Academy of Sciences},
  volume={117},
  number={11},
  pages={5873--5882},
  year={2020},
  publisher={National Acad Sciences},
  doi={10.1073/pnas.1913071117}
}

@article{bisardi2022modeling,
  title={Modeling sequence-space exploration and emergence of epistatic signals in protein evolution},
  author={Bisardi, Matteo and Rodriguez-Rivas, Juan and Zamponi, Francesco and Weigt, Martin},
  journal={Molecular biology and evolution},
  volume={39},
  number={1},
  pages={msab321},
  year={2022},
  publisher={Oxford University Press},
  doi={10.1093/molbev/msab321}
}

@article{alvarez2022novel,
  title={Novel sequence space explored by functional proteins generated through computational evolution-based design},
  author={Alvarez, Sophia and Nartey, Charisse and Mercado, Nicholas and Morcos, Faruck},
  journal={Biophysical Journal},
  volume={121},
  number={3},
  pages={45a},
  year={2022},
  publisher={Elsevier},
  doi={10.1016/j.bpj.2021.11.2476}
}

@article{alvarez2024vivo,
  title={In vivo functional phenotypes from a computational epistatic model of evolution},
  author={Alvarez, Sophia and Nartey, Charisse M and Mercado, Nicholas and de la Paz, Jose Alberto and Huseinbegovic, Tea and Morcos, Faruck},
  journal={Proceedings of the National Academy of Sciences},
  volume={121},
  number={6},
  pages={e2308895121},
  year={2024},
  publisher={National Acad Sciences},
  doi={10.1073/pnas.2308895121}
}

@article{morcos2014coevolutionary,
  title={Coevolutionary information, protein folding landscapes, and the thermodynamics of natural selection},
  author={Morcos, Faruck and Schafer, Nicholas P and Cheng, Ryan R and Onuchic, Jos{\'e} N and Wolynes, Peter G},
  journal={Proceedings of the National Academy of Sciences},
  volume={111},
  number={34},
  pages={12408--12413},
  year={2014},
  publisher={National Acad Sciences},
  doi={10.1073/pnas.1413575111}
}

@article{rodriguez2022epistatic,
  title={Epistatic models predict mutable sites in SARS-CoV-2 proteins and epitopes},
  author={Rodriguez-Rivas, Juan and Croce, Giancarlo and Muscat, Maureen and Weigt, Martin},
  journal={Proceedings of the National Academy of Sciences},
  volume={119},
  number={4},
  pages={e2113118119},
  year={2022},
  publisher={National Acad Sciences},
  doi={10.1073/pnas.2113118119}
}

@article{vigue2022deciphering,
  title={Deciphering polymorphism in 61,157 Escherichia coli genomes via epistatic sequence landscapes},
  author={Vigu{\'e}, Lucile and Croce, Giancarlo and Petitjean, Marie and Rupp{\'e}, Etienne and Tenaillon, Olivier and Weigt, Martin},
  journal={Nature Communications},
  volume={13},
  number={1},
  pages={4030},
  year={2022},
  publisher={Nature Publishing Group UK London},
  doi={10.1038/s41467-022-31643-3}
}

@article{Rossi_2025,
doi = {10.1088/1361-6633/adea92},
url = {https://doi.org/10.1088/1361-6633/adea92},
year = {2025},
month = {jul},
publisher = {IOP Publishing},
volume = {88},
number = {7},
pages = {078102},
author = {Rossi, Saverio and Di Bari, Leonardo and Weigt, Martin and Zamponi, Francesco},
title = {Fluctuations and the limit of predictability in protein evolution},
journal = {Reports on Progress in Physics}
}

@article{marks2011protein,
    doi = {10.1371/journal.pone.0028766},
    author = {Marks, Debora S. AND Colwell, Lucy J. AND Sheridan, Robert AND Hopf, Thomas A. AND Pagnani, Andrea AND Zecchina, Riccardo AND Sander, Chris},
    journal = {PLOS ONE},
    publisher = {Public Library of Science},
    title = {Protein 3D Structure Computed from Evolutionary Sequence Variation},
    year = {2011},
    month = {12},
    volume = {6},
    pages = {1-20},
    number = {12},
}

@Inbook{Calvanese2025generating,
author="Calvanese, Francesco
and Weigt, Martin
and Nghe, Philippe",
editor="Churkin, Alexander
and Barash, Danny",
title="Generating Artificial Ribozymes Using Sparse Coevolutionary Models",
bookTitle="RNA Design: Methods and Protocols",
year="2025",
publisher="Springer US",
address="New York, NY",
pages="217--228",
isbn="978-1-0716-4079-1",
doi="10.1007/978-1-0716-4079-1_15"
}

@article{Lambert2024expanding,
  title={Exploring the space of self-reproducing ribozymes using generative models},
  author={Lambert, Camille N. and Opuu, Vaitea and Calvanese, Francesco and Pavlinova, Polina and Zamponi, Francesco and Hayden, Eric J. and Weigt, Martin and Smerlak, Matteo and Nghe, Philippe},
  journal={Nature communications},
  volume={16},
  number={1},
  pages={7836},
  year={2025},
  publisher={Nature Publishing Group UK London},
  doi={10.1038/s41467-025-63151-5}
}

@article{supekarAncestralSequenceReconstruction2026,
  title = {Ancestral {{Sequence Reconstruction Reveals New Functional Fluorinases}} and {{Mechanistic Insights}} into {{Enzymatic Fluorination}}},
  author = {Supekar, Shreyas and Yeo, Wan Lin and Tiong, Elaine and Rizal, Juliana and Ang, Ee Lui and Wong, Fong Tian and Lim, Yee Hwee and Fan, Hao},
  year = 2026,
  month = feb,
  journal = {Chemical Communications},
  publisher = {The Royal Society of Chemistry},
  issn = {1364-548X},
  doi = {10.1039/D5CC06378G},
  urldate = {2026-02-09},
  langid = {english}
}

@article{chernyavskayaAncestralIntronicSplicing2026,
  title = {Ancestral Intronic Splicing Regulatory Elements in the {{SCN$\alpha$}} Gene Family},
  author = {Chernyavskaya, Ekaterina and Vorobeva, Margarita and Spirin, Sergei A. and Skvortsov, Dmitry A. and Pervouchine, Dmitri},
  year = 2026,
  month = feb,
  journal = {RNA},
  pages = {rna.080730.125},
  address = {New York, N.Y.},
  issn = {1469-9001},
  doi = {10.1261/rna.080730.125},
  langid = {english},
  pmid = {41714109}
}

@article{zhaoAncestralSequenceReconstruction2026,
  title = {Ancestral {{Sequence Reconstruction}} for {{Novel Bifunctional Glutathione Synthetase}} with {{Enhanced Thermostability}} and {{Catalytic Efficiency}}},
  author = {Zhao, Jieru and Wang, Binhao and Di, Junhua and Zhou, Jieyu and Dong, Jinjun and Ni, Ye and Han, Ruizhi},
  year = 2026,
  month = jan,
  journal = {Foods},
  volume = {15},
  number = {2},
  pages = {309},
  address = {Basel, Switzerland},
  issn = {2304-8158},
  doi = {10.3390/foods15020309},
  langid = {english},
  pmid = {41596908}
}

@article{chantreauAsymmetricalDiversificationReceptorligand2019,
  title = {Asymmetrical Diversification of the Receptor-Ligand Interaction Controlling Self-Incompatibility in {{Arabidopsis}}},
  author = {Chantreau, Maxime and Poux, C{\'e}line and Lensink, Marc F and Brysbaert, Guillaume and Vekemans, Xavier and Castric, Vincent},
  editor = {McCormick, Sheila and Hardtke, Christian S and McCormick, Sheila and Barton, Nick H},
  year = 2019,
  month = nov,
  journal = {eLife},
  volume = {8},
  pages = {e50253},
  publisher = {eLife Sciences Publications, Ltd},
  issn = {2050-084X},
  doi = {10.7554/eLife.50253},
  urldate = {2026-04-08}
}

@article{williamsAssessingAccuracyAncestral2006,
  title = {Assessing the {{Accuracy}} of {{Ancestral Protein Reconstruction Methods}}},
  author = {Williams, Paul D. and Pollock, David D. and Blackburne, Benjamin P. and Goldstein, Richard A.},
  year = 2006,
  month = jun,
  journal = {PLOS Computational Biology},
  volume = {2},
  number = {6},
  pages = {e69},
  publisher = {Public Library of Science},
  issn = {1553-7358},
  doi = {10.1371/journal.pcbi.0020069},
  urldate = {2026-04-08},
  langid = {english}
}

@article{hortaEffectPhylogeneticCorrelations2021,
  title = {On the Effect of Phylogenetic Correlations in Coevolution-Based Contact Prediction in Proteins},
  author = {Horta, Edwin Rodriguez and Weigt, Martin},
  year = 2021,
  month = may,
  journal = {PLOS Computational Biology},
  volume = {17},
  number = {5},
  pages = {e1008957},
  publisher = {Public Library of Science},
  issn = {1553-7358},
  doi = {10.1371/journal.pcbi.1008957},
  urldate = {2026-04-13},
  langid = {english}
}

@book{vigueDecipheringPolymorphism611572022,
  title = {Deciphering Polymorphism in 61,157 {{Escherichia}} Coli Genomes via Epistatic Sequence Landscapes},
  author = {Vigue, Lucile and Croce, Giancarlo and Petitjean, Marie and Rupp{\'e}, Etienne and Tenaillon, Olivier and Weigt, Martin},
  year = 2022,
  month = jan,
  doi = {10.1101/2022.01.21.477185}
}

@article{leGascuelModelingProteinEvolution2012,
  title = {Modeling Protein Evolution with Several Amino Acid Replacement Matrices Depending on Site Rates},
  author = {Le, Si Quang and Dang, Cuong Cao and Gascuel, Olivier},
  year = 2012,
  month = oct,
  journal = {Molecular Biology and Evolution},
  volume = {29},
  number = {10},
  pages = {2921--2936},
  issn = {1537-1719},
  doi = {10.1093/molbev/mss112},
  langid = {english},
  pmid = {22491036}
}

\end{document}